\documentclass{article}
\usepackage{booktabs}%
\usepackage{amsmath} 
\usepackage{bookmark}

\usepackage{graphics}
\usepackage{graphicx}
\usepackage{xcolor}
\usepackage{bbding}
\usepackage{pifont}
\usepackage{wasysym}
\usepackage{amssymb}
\usepackage[a4paper, total={6.5in, 9.1in}]{geometry}
\usepackage[labelfont=bf, font=small]{caption}

\usepackage{url} 

\begin{document}

\title{\textbf{Social inequalities that matter for contact patterns, vaccination, and the spread of epidemics}}

\author{Adriana Manna$^1$ \and J\'ulia Koltai$^{2,3,4}$\and M\'arton Karsai$^{1,4,5*}$}

\date{%
\small \emph{
    $^1$Central European University, Quellenstraße 51, 1100 Vienna, Austria\\%
    $^2$Computational Social Science - Research Center for Educational and Network Studies, Centre for Social Sciences, Tóth Kálmánutca 4,Budapest, 1097, Hungary\\%
    $^3$Department of Social Research Methodology, Faculty of Social Sciences, Eötvös Loránd University, Pázmány Péter s étány 1/A, Budapest, 1117, Hungary.\\%
    $^4$ National Laboratory for Health Security, Hungary.\\%
    $^5$ Rényi Institute of Mathematics, Reáltanodautca 13-15, Budapest, 1053, Hungary.\\%
    $^*$Corresponding author: mkarsai@ceu.edu}
}

\maketitle

\begin{abstract}
Individuals socio-demographic and economic characteristics crucially shape the spread of an epidemic by largely determining the exposure level to the virus and the severity of the disease for those who got infected. While the complex interplay between individual characteristics and epidemic dynamics is widely recognized, traditional mathematical models often overlook these factors. In this study, we examine two important aspects of human behavior relevant to epidemics: contact patterns and vaccination uptake. Using data collected during the Covid-19 pandemic in Hungary, we first identify the dimensions along which individuals exhibit the greatest variation in their contact patterns and vaccination attitudes. We find that generally privileged groups of the population have higher number of contact and a higher vaccination uptake with respect to disadvantaged groups. Subsequently, we propose a data-driven epidemiological model that incorporates these behavioral differences. Finally, we apply our model to analyze the fourth wave of Covid-19 in Hungary, providing valuable insights into real-world scenarios. By bridging the gap between individual characteristics and epidemic spread, our research contributes to a more comprehensive understanding of disease dynamics and informs effective public health strategies.
\end{abstract}

\textbf{keywords:} social inequalities, epidemics, human behaviour, mathematical models, contact matrices

\section{Introduction}\label{sec_intro}

Individuals' socio-demographic and economic characteristics are among the most significant factors that shape the dynamics of epidemic spreading processes. They not only influence the epidemic outcome in the hosting population but largely determine the course and severity of the disease for those who got infected \cite{marmot2008closing}. There is a widespread agreement that pandemics disproportionately affect certain population groups rather than others \cite{mamelund2021association,kikuti2015spatial,mena2021, Burstrom2020, paul2021socio,zhao_harris_ellis_pebody_2015}. Health-related inequalities in the burden of an epidemic partly arise from differences in the level of exposure to viruses and bacteria. These are related to differences in social interactions, mobility patterns and work-related conditions, which are aggravated by disparities in the ability to be compliant with non-pharmaceutical interventions (NPIs), such as self-isolation, home-office and avoiding crowded places \cite{Gozzi2020,Jay2020,valdano2021highlighting,pullano2020evaluating,bonaccorsi2020economic}. At the same time, inequalities in the severity and fatality of a disease can be accounted for the heterogeneity in preexisting individual health conditions, protection attitudes and access to medical care, which are themselves related to socio-demographic and economic factors \cite{Sommer2015,Bambra964}.

Although it is widely recognized that socioeconomic inequalities play a crucial role in the transmission dynamics of diseases, traditional mathematical approaches have often overlooked these factors. Indeed, the state-of-the-art framework of modelling infectious diseases incorporates stratification of the population according to age groups \cite{anderson1991infectious}, while discarding other potential relevant heterogeneities between groups of individuals belonging to different socioeconomic strata. They commonly ignore the mechanisms through which these heterogeneities come into play - both directly and indirectly, in the different phases of an epidemic process. In traditional epidemiological models, contact patterns are usually represented in the aggregate form of an \textit{age contact matrix ($C_{ij}$)}, which encodes information on the average number of contacts that individuals of different age groups have with each other \cite{Leung2017,mossong2008social,melegaro2017, Mistry2021, prem2017projecting, grijalva2015household}. Moreover, not only the description of contact patterns is limited to an age structure but also other epidemiological-relevant factors, such as vaccination uptake, infection fatality rates \cite{gozzi2022anatomy} or susceptibility \cite{Zhang2020} are usually described only by considering differences between different age groups. While age is unarguably one of the most important determinant of these characters, the current literature falls short to understand the role of other social, demographics, and economic factors in shaping human behaviour that are relevant to the epidemic spreading. In recent years, researchers have advocated including social aspects in infectious disease modelling, arguing that the epidemic modelling community is lacking a deep understanding of the mechanisms through which the socioeconomic divide translates into heterogeneities in the spread of infectious diseases \cite{tizzoni2022addressing, buckee2021thinking,bedson2021review,zelner2022there}.

\begin{figure}
    \centering
    \includegraphics[scale = 0.47]{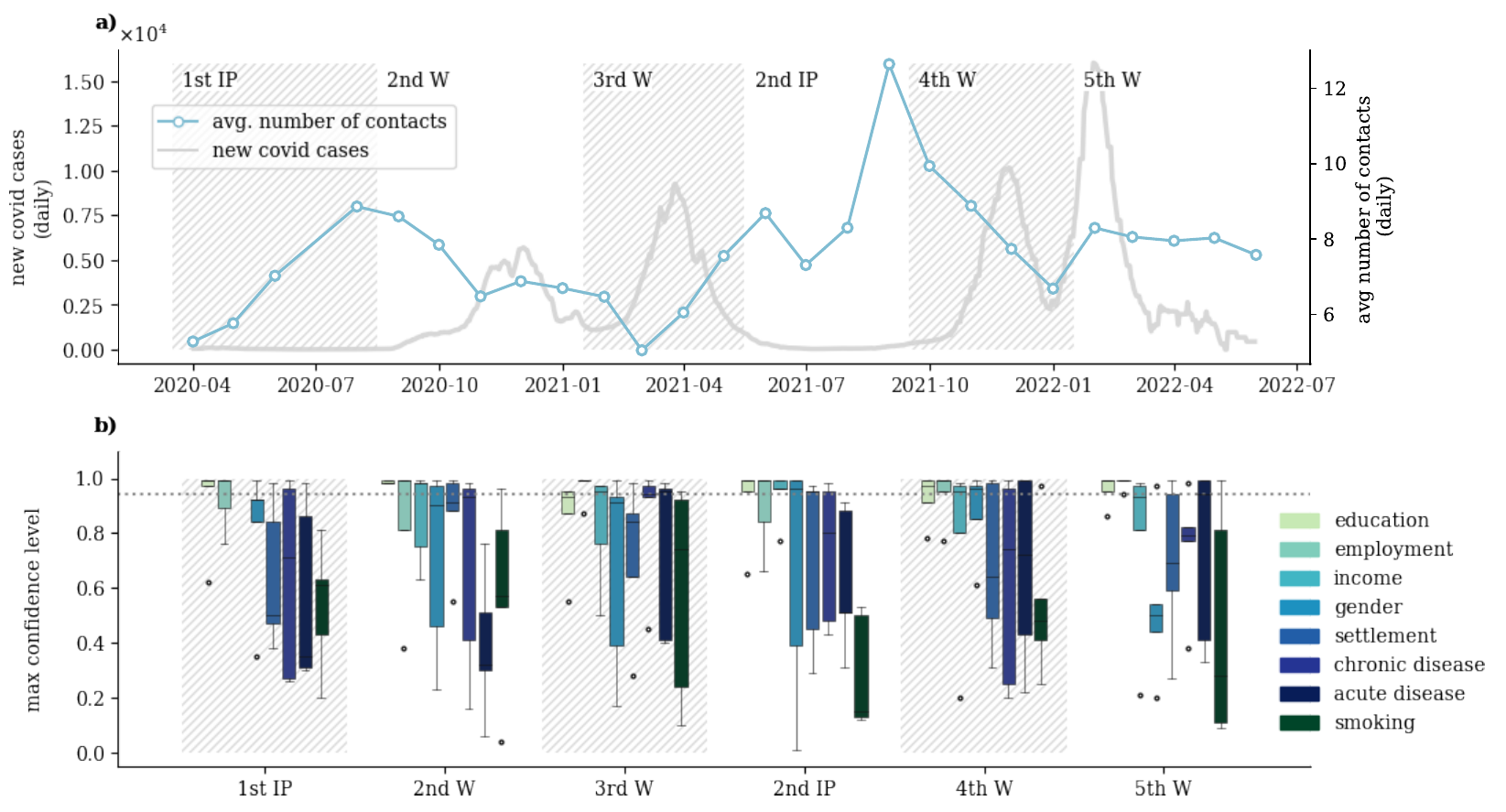}
    \caption{\textbf{(a)} \textit{Left axis}: number of new daily Covid-19 cases in Hungary from 2020/04 to 2022/07. \textit{Right axis}: average number of daily contacts from 2020/04 to 2022/06. The white and grey areas delimit the periods that have been aggregated in the analysis: two interim periods ($IPs$) and four epidemic waves ($W$). \textbf{(b)} Box-plot of the maximum confidence level at which the effect of the different categories of the variable on the total number of contacts becomes significantly different. The dispersion of the box plot refers to the variation of this value over different age groups. Results are shown for education level, employment situation, income level, gender, settlement, chronic disease, acute disease, and smoking behaviour. The higher this value is, the more the variable influences people's number of contacts given their age.}
    \label{fig1}
\end{figure}

With this in mind, we aim to shed light on these mechanisms to address the following interrogatives: which are the most important individual characters and corresponding sub groups of the population that differentiate the most their epidemic-relevant behaviours; and, how these differences translate into epidemiological outcomes? We address these questions by analyzing a large survey dataset coming from the MASZK study carried out in Hungary during the Covid-19 pandemic \cite{karsai2020hungary,koltai2022reconstructing}. This data collects information on individuals' face-to-face interaction patterns in different contexts and other epidemiological-related behavioural patterns and opinions such as travel habits, vaccination attitude, or mask-wearing. The MASZK study consists of $26$ cross-sectional representative surveys carried out in each month between 2020/04 and 2022/06 (for more details on the data see MM). By considering the course of the pandemic in the country, we aggregate the data in six periods covering four epidemic waves (Ws) and two interim periods (IPs) as demonstrated in Fig. \ref{fig1}a.

Throughout this study we are mainly interested in the dynamics and most influential determinants of social contacts, that were recorded in the data as reported proxy interactions between pairs of individuals who spent at least 15 minutes within 2 meters from each other on a given day. Outside of home, we distinguish between two context where social interactions may evolve. We differentiate between \emph{work contacts} that emerge at the workplace (or at school) of respondents (or their minors) and \emph{community contacts} that they evolved elsewhere than home or work. Meanwhile, we do not take into account household contacts in our study as we assume they do not change significantly during the different phases of the pandemic.
Through the analysis of contact patterns, our aim is to show existing significant differences among sub-groups of different socio-demographic characters, 
when accounting for the effect of age. Particularly, we demonstrate that dimensions such as employment situation and education level play a crucial role in determining contact numbers and vaccination uptake during a pandemic. Additionally, by proposing a new data-driven mathematical framework, which explicitly considers further social dimensions, other than age, we analyze the impact of such differences in terms of epidemic outcomes. Finally, by focusing on the Hungarian Covid-19 pandemic scenario, we reveal the unequal impact of the pandemic in terms of individuals belonging to different socio-economic statuses, where we differentiate individuals by their employment situation and income level. Note that although all the models have been completed on each pandemic periods, for the demonstration of our findings we exclusively show results about the $4th$ wave in the main text, while reporting our findings concerning other periods in the Supplementary Information.

\section*{Results}\label{res}

\subsection*{The main determinants of human contact patterns}\label{res2.1}

Human contact patterns represent the routes of infectious disease spreading by shaping the underlining transmission chain among susceptible individuals. During the Covid-19 pandemic, many aspects of human behaviour has experienced drastic interruption in most countries worldwide. This was largely due to the implementation of non-pharmaceutical interventions (NPIs) that were installed to mitigate the spreading and other effects of the pandemic. They aimed at controlling the number of contacts, as well as influencing individual attitudes, to change the ways humans meet and interact with each other \cite{Zhang2020,brankston2021quantifying, trentini2022investigating}. Their effects are evident from  Fig. \ref{fig1}a, where the average number of daily contacts in Hungary are shown. These numbers increase during interim periods (IPs) when the numbers of daily infection cases are low, and decrease during the epidemic waves (Ws) when infection risk is high, this way sensitively reflecting the adaptive behaviour of people throughout the pandemic.

\begin{figure}[ht]
    \centering
    \includegraphics[scale = 0.38]{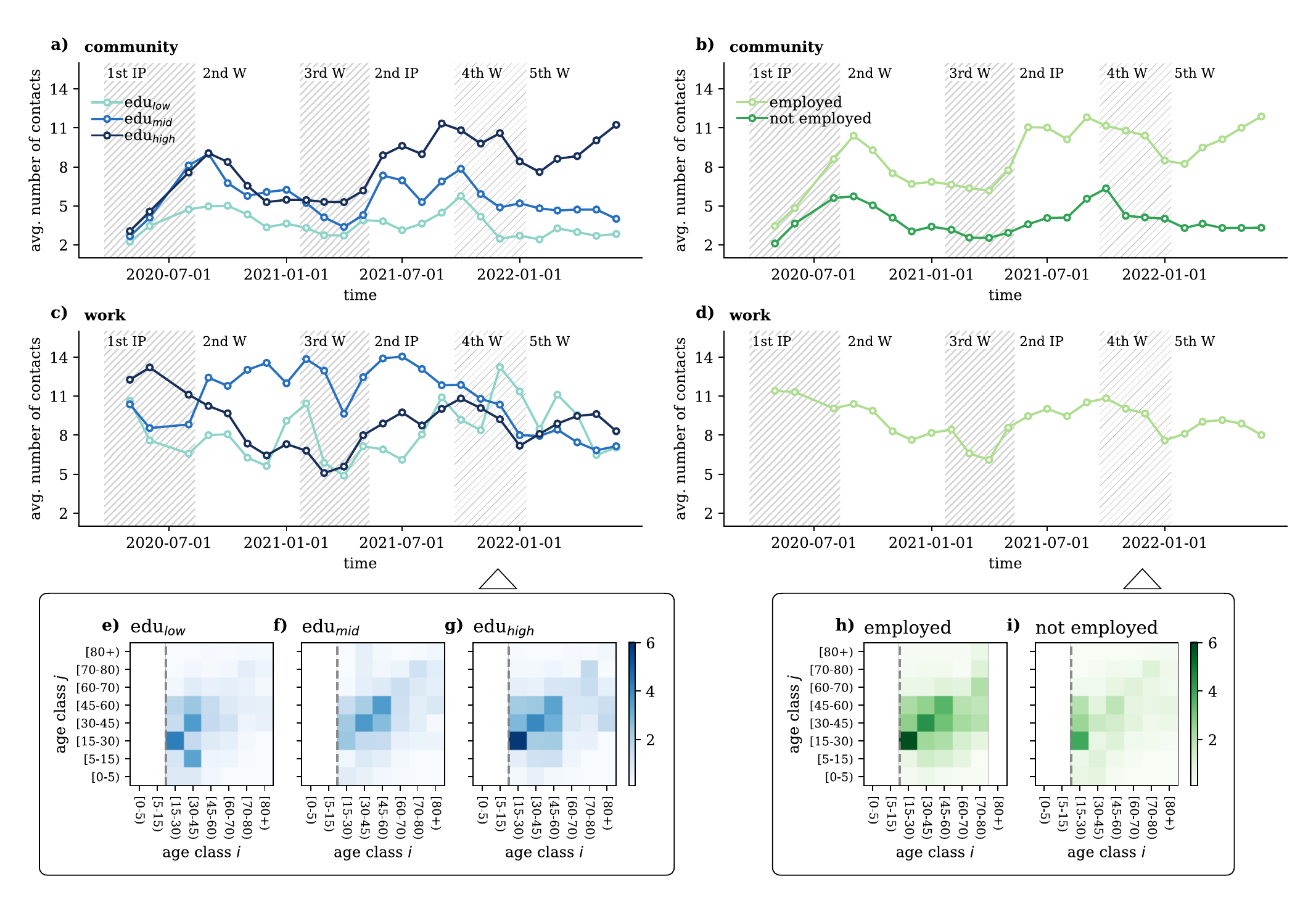}
    \caption{\textbf{(a-b)} Average number of contacts in the community layer (a) by different education levels and (b) employment situation. \textbf{(c-d)} Average number of contacts at workplace (c) by different education levels and (d) for employed people. All curves have been smoothed over the observation periods for better visualisation. \textbf{(e-g)} Decoupled age contact matrices by education level for the 4th epidemic wave. \textbf{(h-i)} Decoupled age contact matrices by employment situation for the 4th wave. These figures depict contact numbers only for adult population [15+), while matrices containing children are shown in SI.}
    \label{fig2}
\end{figure}

Although at the aggregate level these patterns are clear, there are non-trivial disparities at the level of individuals that may result in diverse contact patterns for given sub-groups of the population. To explore these effects, in our statistical analysis, we focus on several socioeconomic dimensions, that, interacting with age, may significantly affect the number of contacts that individuals have. We consider various socio-demographic variables such as individual's education, employment, income, gender, settlement type, actual chronic or acute disease or smoking habits (for more details and definitions see MM and SI). As a first observation, in Fig. \ref{fig1}$b$ we show the distribution of the \textit{maximum confidence level} of the effects of these variables that they had on the number of contacts in interaction with age, during each period (for definition see MM). In these distributions a higher value indicates that a given variable explains better differences in the number of contacts, given the age of individuals. Based on these results employment, education and income level were found to be the three most important dimensions in determining the number of contact. This observation stands if we consider the overall number of contacts including both work and community relationships, and it is true as well if we only consider community contacts (with results shown in the SI).

To further investigate the ways individuals of different characters adapt their number of contacts to the actual epidemiological situation, in Fig \ref{fig2} we show the average number of contacts over time decoupled by education level (Fig \ref{fig2} $a,c$) and employment situation (Fig \ref{fig2} $b,d$) for adult individuals older than 15 years old (for the corresponding plot decoupled by age groups see SI). Results in panel (a) suggest that high and mid-educated individuals have consistently higher number of contacts in the community layer throughout the observed period. In addition, these groups are able to better adapt to the epidemiological situation and NPIs by decreasing their contact numbers during epidemic waves and increasing again during interim periods. At the same time, low-educated individuals maintain a lower number of contacts over time with smaller variation, reflecting their limited social environment and adaption capacities. By looking at the contact dynamics at workplaces it is evident that only highly educated individuals were able to adapt to the epidemiological situation, while the low- and mid-educated people presented similar dynamics and had less flexibility to adapt during the different pandemic periods. Interestingly, mid-educated individuals reported a higher number of contacts at work particularly during the $2nd$ and the $3rd$ epidemic waves.

When we group people by their employment situation, it makes sense to compare groups in the community layer. From Fig \ref{fig2}$b$ it is evident that employed people maintain more contacts even outside of their workplace as compared to not-employed individuals, which is a clear sign of behavioural differences between these two groups. Meanwhile, employed individuals follow somewhat similar contact dynamics as high-educated people (see Fig.~\ref{fig2}$d$, signalling some correlation between these two groups.

From an epidemic modelling perspective, the most convenient way to code interaction patterns between different groups is via contact matrices that represent a social network at an aggregate level. Contact matrices allow models to depart from the homogeneous mixing assumption, i.e. taking all individuals to meet with the same probability. Instead, they allow the introduction of non-homogeneous mixing patterns between different groups, while keeping the model computationally more feasible as compared to contact network based simulations. Conventionally, epidemic models incorporate $C_{i,j}$ age contact matrices that code the average number of contacts between people from different age-groups (for formal definition see MM). Nevertheless, age contact matrices could be further stratified by other socio-demographic characters that influence the contact numbers of individuals. In Fig \ref{fig2}$e-i$ we show the age contact matrices decoupled by education level and employment situation ($C_{di,j}$) for the $4th$ epidemic wave for the adult population (See MM for more details and SI for the corresponding matrices including children). These matrices have been computed by considering \textit{community}, \textit{work} and \textit{household} contacts together. The emerging large differences between these matrices demonstrate clearly that beyond age, the identified variables, i.e. education and employment status, induce significant differences in the contact patterns of people. Although these variables may not be independent of the age of people, the observed distinct patterns suggest more complex mechanisms controlling contact patterns among sub-groups that can be explained by age alone.

\subsubsection*{Beyond age stratification
}\label{res2.2}

We demonstrated that social inequalities significantly influence human contact patterns, thereby shaping the network of proxy social interactions. This is critically important for the propagation of diseases as they determine the transmission chain of an infection spreading among a susceptible population. Consequently, incorporating the contact pattern differences among individuals of different socioeconomic background into epidemiological models is crucial. This could help to understand the unequal spread and uneven burden what an epidemic could impose for the different socio-demographic groups of a society. To this end, we propose a simple mathematical framework based on the extension of a conventional age-structured SEIR compartmental model~\cite{rohani,hethcote2000mathematics}, which we call the \textit{extended SEIR} model. The \textit{conventional SEIR} model assumes that each individual in a population is in one of the mutual exclusive states of Susceptible (S), Exposed (E), Infected (I) or Recovered (I). Transitions of an individual between these states are controlled by rates ($S\xrightarrow{\lambda}E$, $E\xrightarrow{\varepsilon}I$, $I\xrightarrow{\mu}R$) with the $\lambda$ rate influenced by the frequencies of interactions between age groups coded in a $C_{i,j}$ age-contact matrix. The proposed extended model incorporates $C_{\bar{d},i,j}$ age contact matrices instead, that are decoupled along important socio-demographic dimensions $\bar{d}$ to model epidemic spreading in different sub-groups of the population (See MM for further details).

\begin{figure}[ht!]
    \centering
    \includegraphics[scale = 0.44]{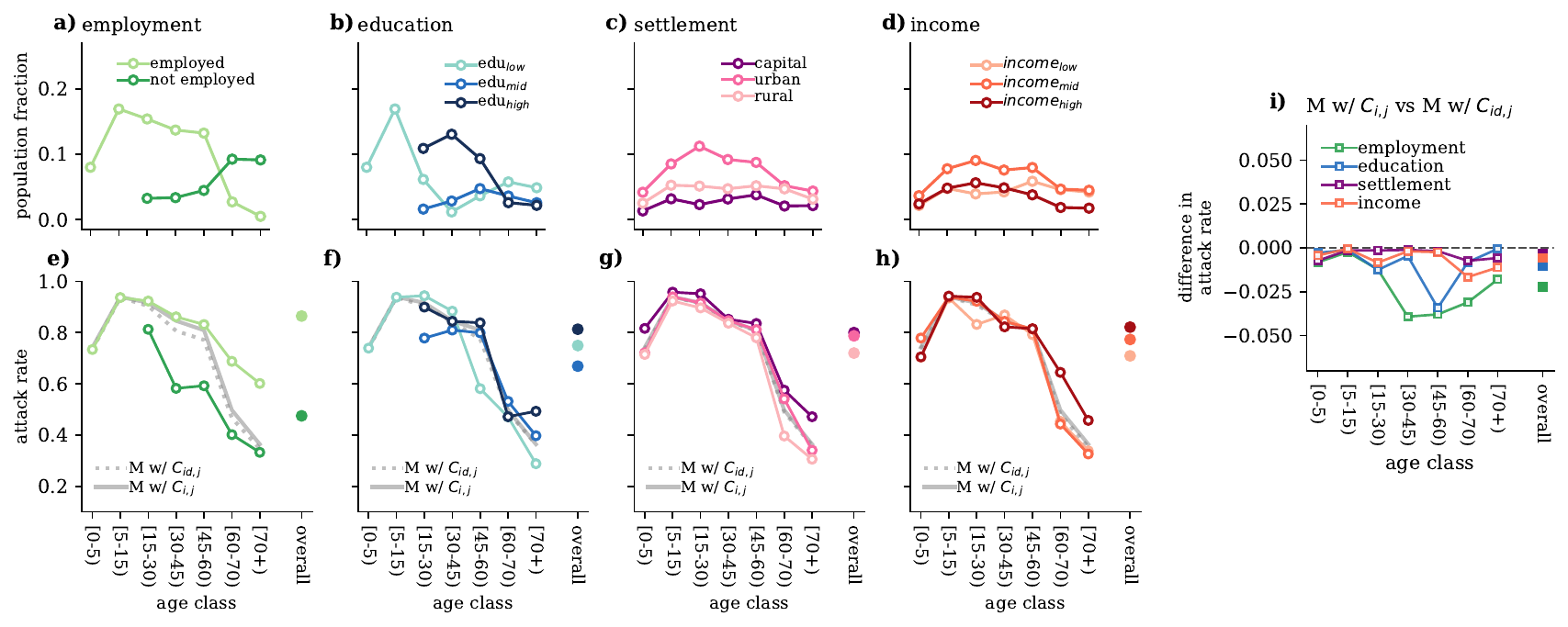}
    \caption{\textbf{(a-d)} Population distribution by age and employment situation $(a)$, education level $(b)$, settlement $(c)$ and income $(d)$. \textbf{(e-h)} Attack rate by age and employment situation $(e)$, education level $(f)$, settlement $(g)$ and income $(h)$ as predicted by the \textit{extended SEIR} (i.e. calculated with both age and the given socioeconomic variable stratification). The grey lines represent the attack rate by age calculated only with age stratification as predicted by \textit{classical SEIR with $C_{i,j}$} (solid lines) and \textit{extended SEIR with $C_{di,j}$} (dotted lines). \textbf{i} Difference in the attack rate by age as predicted by the \textit{ classical} ($Mw/C_{i,j}$) and the \textit{extended} ($Mw/C_{di,j}$) model, when different the dimensions are considered. Results are shown for the $4th$ wave. Epidemiological parameters: $\mu=0.25, \epsilon=0.4$, and $R_0=2.5$. Simulations start with $I_0= 5$ initial infectious seed. Results were computed over 500 simulations with standard deviation smaller than the symbols.}
    \label{fig3}
\end{figure}

Particularly, we analyze the impact of decoupled age-contact matrices along four dimensions: employment situation, education level, settlement, and income level (for exact definitions and possible variable values see MM).
Taking the decoupled contact matrices as input we simulate the spread of infectious disease among an entirely susceptible population using both the \textit{conventional SEIR} and the \textit{extended SEIR} models. Having fixed the epidemiological parameters such as the transmission rates and seeding strategy, other input parameters like the population distributions and contact matrices have been estimated from data as we explain in the SI in more details.

The proposed model allows us to investigate how differences in contact patterns along diverse social groups translate into an unequal burden of the epidemic. To quantify these differences in the epidemic outcome, we measure the attack rate defined as the population wise normalised fraction of individuals who contracted the infection from a given group. To follow the distribution of the people along the investigated dimensions, we show their population fractions in the different age groups in Fig.~\ref{fig3}a-d. Meanwhile, in Fig \ref{fig3}e-h we depict the attack rates calculated using the \textit{extended SEIR} models for different age and socio-demographic groups (and as reference only for age - see grey solid lines). Results are shown for the cases when we decouple each age-group along the four dimensions analyzed. As anticipated by the statistical analysis, employment, and education produce the largest differences between groups in terms of attack rate by age. 
Interestingly, the group of employed people happened to be the most infected group in all age groups, while mid and high educated individuals are more infected among those, who are 45-60 years old. When decoupling age contact matrices by settlement and income, although differences appear smaller between groups, high-income individuals and the ones living in the capital are more infected, particularly elderly ones with age 60+. These modelling results suggest the interesting overall conclusion that privileged people of the population report a higher attack rate, thus they are typically the most infected group relative to their population size.

These results also demonstrate that the \textit{extended SEIR} model is able to capture differences introduced by the considered socio-demographic variable and, in this way, to model the epidemic impact on the different groups of the population. These differences are also visible at the population level. In Fig \ref{fig3}\textit{i} we show the differences between the attack rates predicted by the \textit{conventional} and the\textit{extended SEIR} models, for each age group and overall too. It is evident from these results that models using contacts only stratified by age may underestimate (negative difference) the size of the epidemic in different age groups or in the whole population. For example, our simulations based on data from the $4th$ wave demonstrate that the \textit{conventional SEIR} model could predict higher attack rates for each age group with respect to the \textit{extended SEIR}, which considers differences in contact numbers along the employment situation or the education level too.

It is important to highlight that the uneven age distribution within the different subgroups sometimes reduces or annul the effect of difference in the contact patterns when we are computing aggregate quantities at the population level. This explains why, even if there is a significant difference in contact patterns, the difference in the attack rates only spans a small range between the two models. 

\subsection*{Vaccination uptake and contact number differences}

\begin{figure}[ht!]
    \centering
    \includegraphics[scale = 0.5]{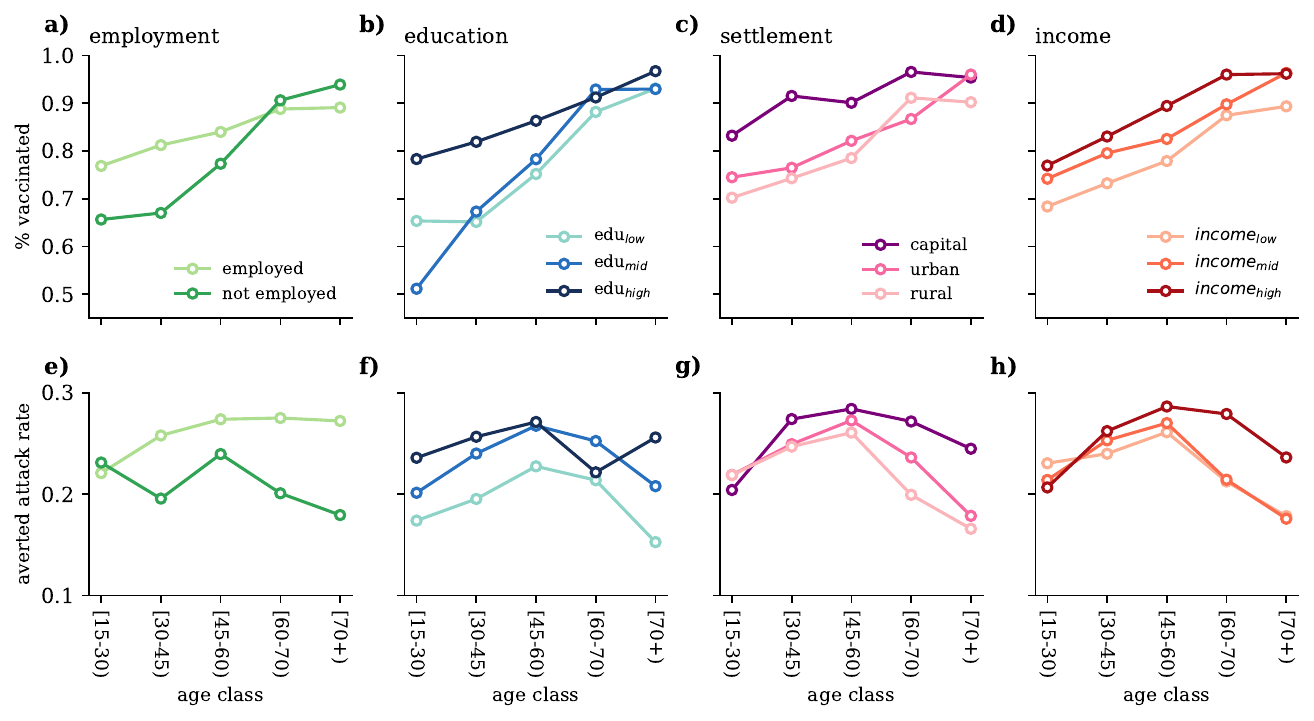}
    \caption{\textbf{(a-d)} Fraction of individuals vaccinated with at least one dose against Covid-19 during the 4th wave, decoupled by age and (a) employment situation, (b) education level, (c) settlement, and (d) income. Panels (e-h) show averted attack rates by (e) age and employment situation, (f) education level, (g) settlement, and (h) income as predicted by \textit{extended SEIR}. The model takes into account different rates of vaccination uptake by subgroups of the given variable, compared to the non-vaccination scenario. Epidemiological parameters: $\mu=0.25, \epsilon=0.4$, and $R_0=2.5$. Simulations start with $I_0= 5$ initial infectious seeds. Results were computed over 500 simulations with standard deviation smaller than the symbols.}
    \label{fig4}
\end{figure}

Beyond the crucial role played by the network of face-to-face interactions, individual vaccination attitudes may also substantially affect epidemiological outcomes by decreasing the morbidity and mortality. By applying the same pipeline of statistical analysis as we explained above, we identified the dimensions along which individuals made different decisions in terms of vaccination, given their age. In this case, the interaction with age is particularly important given that the Covid-19 immunization strategy implemented in Hungary followed an age-stratified outreach by prioritizing elderly individuals~\cite{sandor2022covid,cadeddu2022planning}. Interestingly, the statistical analysis in this case indicated income as the most important dimension along which individuals made different vaccination decisions (See SI for the  results of the statistical analysis). Fig \ref{fig4}a-d show the percentage of vaccinated individuals by age and the investigated dimensions, during the 4th wave of the pandemic. Although by the 4th epidemic wave the vaccination saturated in Hungary, the effects of the age-dependent vaccination policy is clearly visible. More strikingly, we find that privileged groups of the population were more likely to get vaccinated against the Covid-19 virus. Convincingly, this observation is valid for all age groups and periods considered in the analysis (See SI for the corresponding figures for the other periods).

To consider these observations, we model the vaccination uptake in the \textit{extendend SEIR} framework. More precisely, we define the probability of getting vaccinated (i.e. immune or recovered from the point of the infection) to be dependent in this case on both the age and the subgroup of the population considered. Using this \textit{extended SEIR} model, we are able to compare the effects of vaccination uptake, while keeping fixed the structure of contacts. Fig \ref{fig4}e-h shows the \textit{averted attack rate} due to vaccination with respect to the non-vaccination scenario. We consider the probability of getting vaccinated along the four different investigated dimensions separately. In all of these scenarios the gain in averted infection is strongly dependent on the subgroup membership. As expected, the groups with higher vaccination uptakes are the ones, which reduce their attack rate the most in the vaccination scenario. However, this pattern is not linear. For example, among individuals aged $60+$, although the not-employed people report a higher vaccination uptake, they are the ones that gain the less in terms of averted infections. This is because these individuals, having a low number of contact, are already protected from exposure to the virus thus they gain less from vaccination. It should be noted that our focus is solely on the number of infected individuals. It is important to acknowledge that alternative conclusions may arise if the number of averted deaths is taken into consideration. In any case, these results clearly show that although several other factors affect the outcome of an epidemic, by neglecting differences in vaccination uptake and the effects of vaccination campaigns among subgroups in modelling, we miss an important determinant, which significantly influence the final outcome of an epidemic.

\subsection*{Stratified modelling of the Hungarian scenario}

To provide an example of how the proposed mathematical framework can be applied to a real case scenario, we model the $4th$ Covid-19 wave in Hungary between 09/2021 and 01/2022. As the statistical analysis showed that employment and income are the most important dimensions along which, respectively, contact patterns and vaccination uptake change the most, here we divide the population into subgroups by considering simultaneously these two additional dimensions other than age. In addition, we introduce a new compartment $D$ to our SEIR model, that represents a dead state that infected individuals may enter with transmission rate $I\xrightarrow{\mu IFR}D$.

To simulate the SEIRD for the $4th$ Covid-19 wave in Hungary we calibrate our model using the Approximate Bayesian Computation (ABC) method ~\cite{minter2019approximate,sunnaaker2013approximate} on the total number of daily deaths from 09/2021 to 01/2022 \cite{kimittud}. Details about the fitting method and calibrated results are summarised in the SI.

\begin{figure}[ht!]
    \centering
    \includegraphics[scale = 0.18]{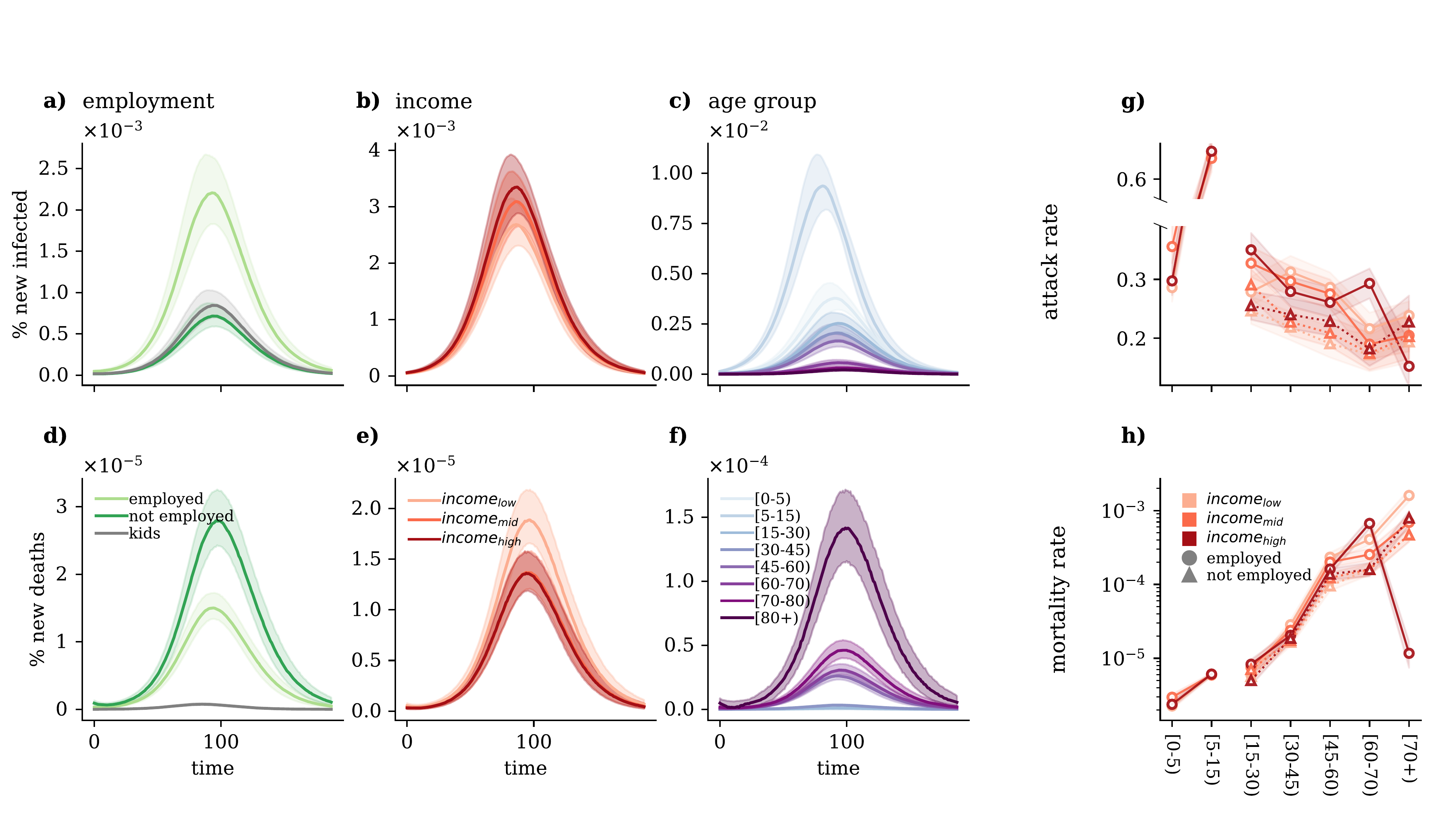}
    \caption{Results from the simulated model for the $4th$ wave in Hungary. The figure shows the median and 95\% confidence interval over 3000 runs. \textbf{(a-c)} Fraction of newly infected by (a) employment situation, (b) income level, and (c) age group. \textbf{(d-f)} Fraction of new deaths by (d) employment situation, (f) income level, and (g) age groups. \textbf{(g)} Attack rate  by age, employment situation and income level. \textbf{(h)} Mortality rate  by age, employment situation and education level. For more details about the parameters of the numerical simulations see SI}.
    \label{fig5}
\end{figure}

The results of the simulated model are presented in Figure \ref{fig5}, which shows the daily fraction of newly infected (panels (a)-(c)) and new dead (panels (d)-(f)) cases for different employment, income, and age groups. As expected, these curves suggests that group of employed people extracted the infection with the highest rate as compared to not employed others. At the same time, in terms of socioeconomic status and age, more affluent and younger people got infected more during the simulated epidemic wave. On the other hand, strikingly the contrary trend is suggested in terms of mortality rates. From the simulations we found that although not-employed, low-income and older individuals appeared with the lowest infection rates, they evolved with the highest mortality rate as compared to other groups. 

Considering that the fatality rate of infected individuals depends on their age, these observation can be largely accounted to the fact that not-employed and low-income individuals are also the oldest one in the population. To further explore the correlation between these dimensions, we examine the attack rates and mortality rates by age groups separately in each of the subgroups of the population stratified by income level and employment status (Fig. \ref{fig5}$g-h$). These results confirm an overall decreasing infection rate by age and that not-employed individuals experience the lowest attack rate in each age group. Further stratification of this group by income level reveals a clear pattern, with high-income people exhibiting a higher infection rate compared to mid-income and low-income individuals. In contrast, the infection pattern among income levels of employed individuals is age-dependent. For young employed individuals in age group 15-30, high and mid-income individuals register a higher share of infections, while among those older than 30, low-income individuals exhibit a higher infection rate compared to mid and high-income individuals. An exception is represented by the age group 60-70, for which the most infected group is still the high income. In terms of mortality we find an increasing trend by age, otherwise we can conclude a similar pattern. Once we account for the correlations among the dimensions of interest and age, our results show that employed people die with a higher rate, while in terms of income, other than the group of age 15-30 and 60-70, the lower income people suffered more death according to our simulations. The extreme decrease of mortality rate for the employed high-income group is due to data sparsity in the survey data, recording only a few data point in this category.

\section*{Discussion}\label{disc}
There are several particular factors that may determine how an infection would turn-out for a given person. Some of them are coded genetically, or determined by physiological conditions, but many of them are environmental and correlate with one's socio-demographic characters. In any society people show uneven patterns along numerous social, demographic, and economic characters, like age, income or employment status. These characteristics not only induce medical disparities between people (as in immunity, overall health conditions, or chronic diseases) but they naturally translate to differences in adaption capacities and other behavioural patterns, letting certain groups more exposed to infection. The simultaneous actions of all these factors lead to observable inequalities in terms of epidemic burden between different groups at the population level.

This study highlights the significant impact that social determinants have on human behaviours that are relevant to epidemic transmission. Specifically, exploiting the data of the \textit{MASZK} study \cite{karsai2020hungary,koltai2022reconstructing}, we show that contact patterns and vaccination uptake are influenced by socioeconomic factors. Our findings suggest that contact patterns are shaped by social factors not only in their absolute values but also in the extent to which they fluctuate in response to extraordinary events, such as a lockdown or curfew interventions. Specifically, our statistical analysis shows that socioeconomic factors such as employment situation and education level played a significant role in determining contact numbers and vaccination uptake during the COVID-19 pandemic in Hungary. Additionally, we find that privileged groups tend to have a higher number of a contact and are the ones able to better adapt to the epidemiological situation and NPIs by adjusting their number of contacts. Contrarily, less privileged groups maintain a lower number of contacts over time with smaller oscillations.  We also find that privileged groups, such as those with higher education, income, and employment status, were more likely to get vaccinated.

We propose a mathematical framework that extends the well-known age-stratified approach to model infectious diseases by explicitly accounting for differences in contact patterns and vaccination uptake for specific sub-groups of the population.
This method allows us to better understand the mechanisms underlying the emergence of inequalities in epidemiological outcomes. Results demonstrate that traditional epidemiological models, that only consider age, could overlook crucial heterogeneities along other social and demographic aspects that may impact the spreading of an epidemic.
By simulating a pandemic period in Hungary, we reveal the unequal health-related impact of the COVID-19 pandemic along individuals belonging to different socioeconomic groups. Although the higher number of contacts translates into higher attack rates for privileged individuals, the age structure and the vaccination decision of such groups translate into lower mortality rates for these individuals, while disadvantaged groups are the one suffering higher mortality. These results are in line with the empirical findings of \cite{oroszi2022characteristics, oroszi2021unequal} for the $2nd$ and $3rd$ Covid-19 waves in Hungary. 

Due to the limitation of the survey collection methodology, contact patterns of individuals can be differentiated only by the characteristics of participants.
Indeed, the only information we know about the contacted peers is their age, while their other characteristics remain unknown. Thus, our extended \textit{SEIR} model can only account for age-contact matrices that are decoupled along other social dimension of the participants (\textit{ego}). In other words, although our model incorporates additional social dimensions, given the sub-group the \textit{ego} belongs to, it still only considers the average number of contacts stratified by the age group of the contacted (\textit{alter}). In order to introduce a generalised matrix~\cite{manna2023generalized} stratified along multiple socio-demographic dimensions of the contactee, we would need information about such dimensions. Such information can be collected via detailed contact diaries~\cite{koltai2022reconstructing}, which are based on the reports of the respondents about the peers, and commonly suffer from recall biased and other limitations.


By shedding light on the complex interplay between social, demographic and economic factors and disease transmission dynamics, our findings underline the need for a new mathematical framework for epidemic modelling that accounts for multidimensional inequalities. This would help us to better understand the socially stratified consequences of an epidemic and to highlight non-negligible inequalities between  different socio-demographic groups. Additionally, incorporating social factors into epidemiological models will provide a valuable tool to design and evaluate targeted NPIs to cope more efficiently with the spread of an infectious disease.

\section{Materials and Methods}\label{method}
\label{sec_method}


\subsection*{Data description}

The data used in this study comes from the MASZK survey study \cite{karsai2020hungary,koltai2022reconstructing}, a large data collection effort on social mixing patterns made during the COVID-19 pandemic. It was carried out in Hungary from April 2020 to July 2022 on a monthly basis. The data was collected via cross-sectional anonymous representative phone surveys using CATI methodology and involved a 1000 large nationally representative sample each month. During the data collection participants were not asked information that could be used for their re-identification. The data collection was fully complying with the actual European and Hungarian privacy data regulations and was approved by the Hungarian National Authority for Data Protection and Freedom of Information~\cite{naih}, and also by the Health Science Council Scientific and Research Ethics Committee (resolution number IV/3073- 1 /2021/EKU).

The primary goal of the data collection effort was to follow how people changed their social contact patterns during the different intervention periods of the pandemic. Relevant to this study, the questionnaires recorded information about the \emph{proxy social contacts}, defined as interactions where the respondent and a peer stayed within 2 meters for more than 15 minutes~\cite{ProxyContactDef} at least one of them not wearing mask. Approximate contact numbers were recorded between the respondents and their peers from different age groups of 0–4, 5–14, 15–29, 30–44, 45–59, 60–69, 70–79, and 80+. Contact number data about underage children were collected by asking legal guardians to estimate daily contact patterns.

Beyond information on contacts before and during the pandemic, the MASZK dataset provided us with an extensive set of information on \textit{social-demographic characteristics} (gender, education level etc.), \textit{health condition} (chronic and acute illness etc.), \textit{financial and working situation} (income, employment status, home office etc.), and \textit{attitude towards Covid-19 related measures and recommendations} (attitude towards vaccination, mask-wearing etc.) of the participants. In order to study different stages of the pandemic, we consider six epidemiological periods including three epidemic waves (\textit{Ws}) and three interim periods (\textit{IPs}) (see Fig \ref{fig1}a).

On the collected data a multi-step, proportionally stratified, probabilistic sampling procedure was elaborated and implemented by the survey research company using a database that contained both landline and mobile phone numbers. The survey response rate was 49 percent, which is expressly higher than the average response rate (being between 15-20 percent) of telephone surveys in Hungary. The sample is representative for the Hungarian population aged 18 or older by gender, age, education and domicile. Sampling errors were corrected using iterative proportional post-stratification weights. After data collection, only the anonymised and hashed data was shared with people involved in the project after signing non-disclosure agreements.

\subsection*{Sociodemographic dimensions}

The sociodemographic dimensions that we analyze are the following: \textit{(i) education level}, which can have three possible levels: low, mid and high; \textit{(ii) employment situation}, which can be either employed or not-working, including students and retirees individuals; \textit{(iii) income} can have three possible levels: low, mid and high; \textit{(iv) gender} refers to the biological gender and can be either female or male; \textit{(v) settlement}, which refers to the area where individual live and can be either capital, rural or urban; \textit{(vi) chronic disease} is a boolean dimension indicating if an individual is affected by any chronic disease; \textit{(vii) acute disease} is a boolean dimension indicating if an individual is affected by any acute disease; and \textit{(viii) smoking} is a boolean dimension indicating if an individual is a smoker or not. A detailed explenation of these variables is provided in the SI.

\subsection*{Statistical analysis}

In order to build an epidemiological model that explicitly takes into account social inequalities we need to identify, which are the main dimensions, that interacting with age affect contact patterns the most. To identify these dimensions, we model the expected number of contacts of respondents $i$ using a negative binomial regression \cite{feehan2021quantifying, mossong2008social} as defined in Eq.~\ref{stat_model}:
\begin{equation}
    \mu_i = \alpha + \beta_1 age\_group_i + \beta_2 X_i + \beta_3 age\_group_i* X_i + \epsilon_i,
\label{stat_model}
\end{equation}
where $age\_group_i$ is the age class of $i$; $X_i$ is the variable of interest (e.g., education, income  etc.), $age\_group_i*X_i$  is the interaction term of age group and the variable of interest, and $\epsilon_i$ is the error term. Given $\mu_i$, we define $\lambda_i=exp(\mu_i)$ to be the expected number of contacts for respondent $i$. Then we model the reported number of contacts for respondent $i$, $y_i$, as
\begin{equation}
   y_i \sim Neg-Bin(\lambda_i, \phi)
\end{equation}
where $ \phi \in [1, \infty)$  is a shape parameter that is inversely related to over-dispersion: the higher $\phi$ is estimated to be, the closest $y_i$’s distribution is to a Poisson distribution with rate parameter $\lambda_i$.

We build a model \ref{stat_model} for each variable of interest ($X$). Particularly, the interaction term between $age\_group_i$ and the variable of interest allows us to examine, whether there are differences in the effect of $X_i$ on the number of contacts in the different age groups. To be able to provide a meaningful description of the interactions, we analyse the average marginal effect \textit{(AME)} \cite{brambor2006understanding,mood2010logistic,allison1999comparing} of $X_i$ on the number of contacts for different age groups, defined as:
\begin{equation}
\label{ame}
\begin{array}{cc} 
AME_{X_i} = \frac{1}{n} \sum_{i=1}^{n} \frac{\partial \mu_i}{\partial age\_group_i}\\ 
\\
 \frac{\partial \mu_i}{\partial age\_group_i} = \beta_1 + \beta_3 X_i.
\end{array}
\end{equation}

\begin{figure}
    \centering
    \includegraphics[scale = 0.55]{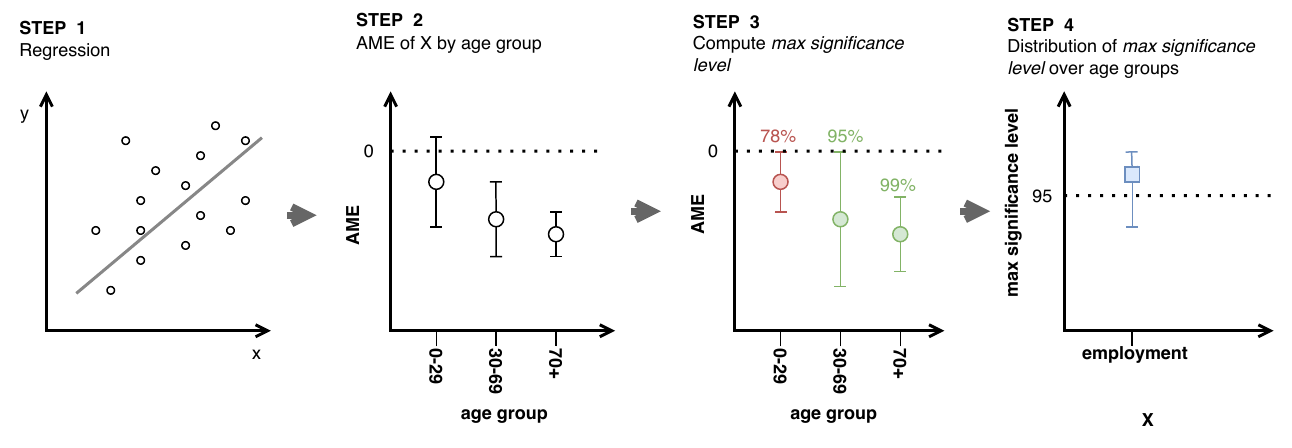}
    \caption{Pipeline of the statistical analysis to compute the distribution of the \textit{max confidence level} over age groups for each of the variables analyzed. The figure is made considering as an illustrative example the variable \textit{employment}.}
    \label{stat_method_fig}
\end{figure}

Working mostly with categorical variables (e.g., education level or employment situation), we can calculate different AMEs for all categories of the categorical variables in each $age\_group$. For each age group, we consider the maximum confidence level, at which the $AME$s of two categories of the categorical variable show a statistically significant difference. Finally, to summarize the results, we consider the median of these maximum confidence levels. Therefore, we have one value for each variable of interest, which characterise the strength of the interaction between age and the variable of interest on the number of contacts. By following this procedure for each of the variables of interest, we are able to rank the variables according to their importance in driving differences in contact patterns additionally to age, in different periods of the Covid-19 pandemic. The pipeline of these analyses are illustrated in Fig.~\ref{stat_method_fig}.

Following the same methodology we investigate the dimensions that - in interaction with age - affect the most the probability of getting vaccinated against Covid-19. In this case, we model this probability using a logistic regression model instead of a negative binomial, as the dependent variable was binary and not a count one. In the Supplementary Information we provide the details of this models and we report their relative results.

\subsection*{Decoupled contact matrices}

Conventionally, to compute age-contact matrices $C_{ij}$ we divide a population into sub-groups according to their age and calculate the average number of contacts that individuals in age class $i$ have with individuals in age class $j$~\cite{koltai2022reconstructing}. Here, instead, we further stratify individuals from each age class $i$ according to various dimensions, like employment status, settlement or education level. 


In detail, we decouple the conventional age contact matrix $C_{ij}$ into $D$ number of matrices, one for each of the sub-groups of the dimension that we want to take into account. More precisely, let $\bar{d}$ be the sub-group of the dimension considered and let $\bar{d} \in 1,..., D$. We can write
\begin{equation}
    C_{\bar{d}i,j} = CT_{\bar{d}i,j}/N_{\bar{d}i},
\end{equation}
where $CT_{\bar{d}i,j}$ is the total number of contacts that individuals of age class $i$ and belonging to sub-group $\bar{d}$ have with individuals in age class $j$, regardless of the sub-group, to which the contacted individuals belong; and $N_{\bar{d}i}$ is the total number of individuals in age class $i$ and sub-group $\bar{d}$.

For example, to differentiate between \textit{employed} and \textit{not-employed} individuals, we compute two age contact matrices: $C_{employed,i,j }$ and $C_{not-employed,i,j}$. Note that here we are considering each of these dimensions (e.g., employment, education level, settlement) separately and only include one dimension additionally to age. However, this framework can be extended to any number of dimensions considered simultaneously, in this case, the length of the $\bar{d}$ vector will correspond to the number of combinations of the levels of the dimensions considered.

\subsection*{The epidemiological model}

In order to investigate the effect of the decoupled contact matrices on the dynamic of infectious disease transmission, we propose a simple mathematical framework as an extension of the conventional age-structured SEIRD compartmental model \cite{rohani,hethcote2000mathematics}. 

The conventional SEIRD model is defined on a population where individuals are assigned to five compartments based on their actual state: susceptible ($S$), exposed ($E$), infected ($I$), recovered ($R$) and dead ($D$). The model further defines the transition rates of individuals from one compartment to another by incorporating for each age class a given force of infection, which includes the average number of contacts with all the other age classes. The model proposed here extends this definition by taking into account not only the age structure of the contacts in the population but also their differences along a set of other dimensions $\bar{d}$, such as education level, income level and employment situation.

The model can be described by a set of ordinary coupled differential as presented in Eq. \ref{eq_SEIRD}:

\begin{equation}
\label{eq_SEIRD}
\begin{array}{lcl} 
\dot{S}_{\bar{d},i} & = & -\lambda_{\bar{d},i} S_{\bar{d},i} \\ 
\dot{E}_{\bar{d},i} & = & \lambda_{\bar{d},i} S_{\bar{d},i} -\epsilon E_{\bar{d},i} \\ 
\dot{I}_{\bar{d},i} & = & \epsilon E_{\bar{d},i} - \mu I_{\bar{d},i} \\
\dot{R}_{\bar{d},i} & = &  \mu (1-IFR_i) I_{\bar{d},i} \\
\dot{D}_{\bar{d},i} & = &  \mu IFR_i I_{\bar{d},i}. \\ 
\end{array}
\end{equation}

Here $i$ indicates the age group of the ego, $j$ indicates the age group of the peer, $\bar{d}$ represents a vector of dimensions to which the ego belongs, $\beta$ is the probability of transmission given a contact,$\epsilon$ is the rate at which individuals become infectious, $\mu$ is the recovery rate, $IFR$ is the infection fatality rate, and $C_{\bar{d}}$ is the age contact matrix corresponding to dimensions $\bar{d}$. 

In this equation system we rely on the concept and of force of infection that is defined as:
\begin{equation}
\begin{array}{lcl} 
\lambda_{\bar{d},i}(t)=  \beta \sum_{j} \frac{C_i{\bar{d},j}}{N_{j}} {I_{j}}
\end{array}
\end{equation}
Further we rely on the definition of the \emph{infection fatality rate} ($IFR_i$), that is defined as fraction of infected individuals that died. See Supplementary Information for the details on the implementation of the numerical simulations.

\section*{Acknowledgment}
The authors gratefully thank to Alessandro Vespignani, Eszter Bok\'ani, Alessia Melegaro and Filippo Trentini for useful discussions. A.M. and M.K. were supported by the Accelnet-Multinet NSF grant. J.K. and M.K. acknowledges funding from the National Laboratory for Health Security (RRF-2.3.1-21-2022-00006). M.K. acknowledges support from the ANR project DATAREDUX (ANR-19-CE46-0008); the SoBigData++ H2020-871042; the EMOMAP CIVICA projects.

\clearpage
\section*{\LARGE{\textbf{Supplementary Informations}}}

\section*{Data pre-processing}
All the analysis on the number of contacts have been performed after having delete the outliers at the $99\%$ percentile with respect to the period of interest. 

All the results presented in this work have been computed by accounting each participant according to it's representative weight. The weight has been provided by the survey company as described in the MM section of the main text. 

\section*{Socio-demographic dimensions}
The MASZK dataset provided us with an extensive set of information on \textit{social-demographic characteristics} of the participants. In this section we provide a detailed explanation of all the variables used in this work. 
\begin{itemize}
    \item  \textit{Education level} which can have three possible levels: low, mid and high. We considered low educated individuals those with any primary school degrees, mid educated those holding any diploma or certificate from professional schools, and high educate those with an university education or above (eg. BSc, MSc, PhD).
    \item  \textit{Employment situation} which can be either employed or not-employed. In the not-employed category we include students and retirees individuals. 
    \item  \textit{Income level} can have three possible levels: low, mid and high. In particular, individuals were asked to report their perceived income with respect to the average using a scale from 1 to 10. We consider as low income individual those that answered from 1 to 4, mid income those who answered 5 or 6 and high income those that answered 7 or above.
    \item  \textit{Gender} refers to the biological gender and can be either female or male.
    \item  \textit{Settlement} which refer to the area where individual live and can be either capital, rural or urban.
    \item  \textit{Chronic disease} is a boolean dimension indicating if an individual is affected by any chronic disease or not. 
    \item  \textit{ Acute disease} is a boolean dimension indicating if an individual is affected by any acute disease or not. 
    \item  \textit{Smoking} is a boolean dimension indicating if an individual is a smoker or not. We consider as \textit{non smoker} individuals that declared them self as not-smoker or that stopped, while we consider smokers individuals that smoke frequently or occasionally. 
\end{itemize}

\section*{Statistical Model}
In this section we provide all the results of the additional statistical analysis that we performed in order to support the findings presented in the Result section of the main text. 

\subsection*{Model on contacts}
\subsubsection*{Accounting for the high number of zeros}
Due to the implementation of NPIs (lockdowns and curfews) throughout the pandemic, people were forced, when possible, to reduce or completely reset their number of contacts. Thus, when analysing the distribution of the number of contacts we found a high presence of zeros.

To test the robustness of the results, coming from the negative binomial regression model ($nb$), to this zero-inflated mechanism we implemented two additional models:
\begin{enumerate}
    \item After having excluded the observations where the number of contacts were zero we re-run the \textit{negative binomial regression model} on the non-zero number of contacts ($nb_{n contacts>0}$).
\begin{equation}
    \mu_i = \alpha + \beta_1 age\_group_i + \beta_2 X_i + \beta_3 age\_group_i* X_i + \epsilon_i
\label{stat_model_nbzi}
\end{equation}

    \item We modelled the probability to have at least a contact using a \textit{logistic regression model} ($logit$).

\begin{equation}
    \log P_i(any contact=1) = \alpha + \beta_1 age\_group_i + \beta_2 X_i + \beta_3 age\_group_i* X_i + \epsilon_i
\label{stat_model_logitzi}
\end{equation}
\end{enumerate}

where $age\_group_i$ is the age class of $i$; $X_i$ is the variable of interest (e.g., education, income  etc.), $age\_group_i* X_i$  is the interaction term of age group and the variable of interest, and $\epsilon_i$ is the error term. 
In model \ref{stat_model_nbzi} given $\mu_i$, we define $\lambda_i=exp(\mu_i)$ to be the expected number of contacts for respondent $i$. While in model \ref{stat_model_logitzi} $P_i(any contact=1)$ indicate the probability for respondent $i$ to have at least one contact.

\begin{figure}[ht!]
  \centering
  \includegraphics[scale =0.55]{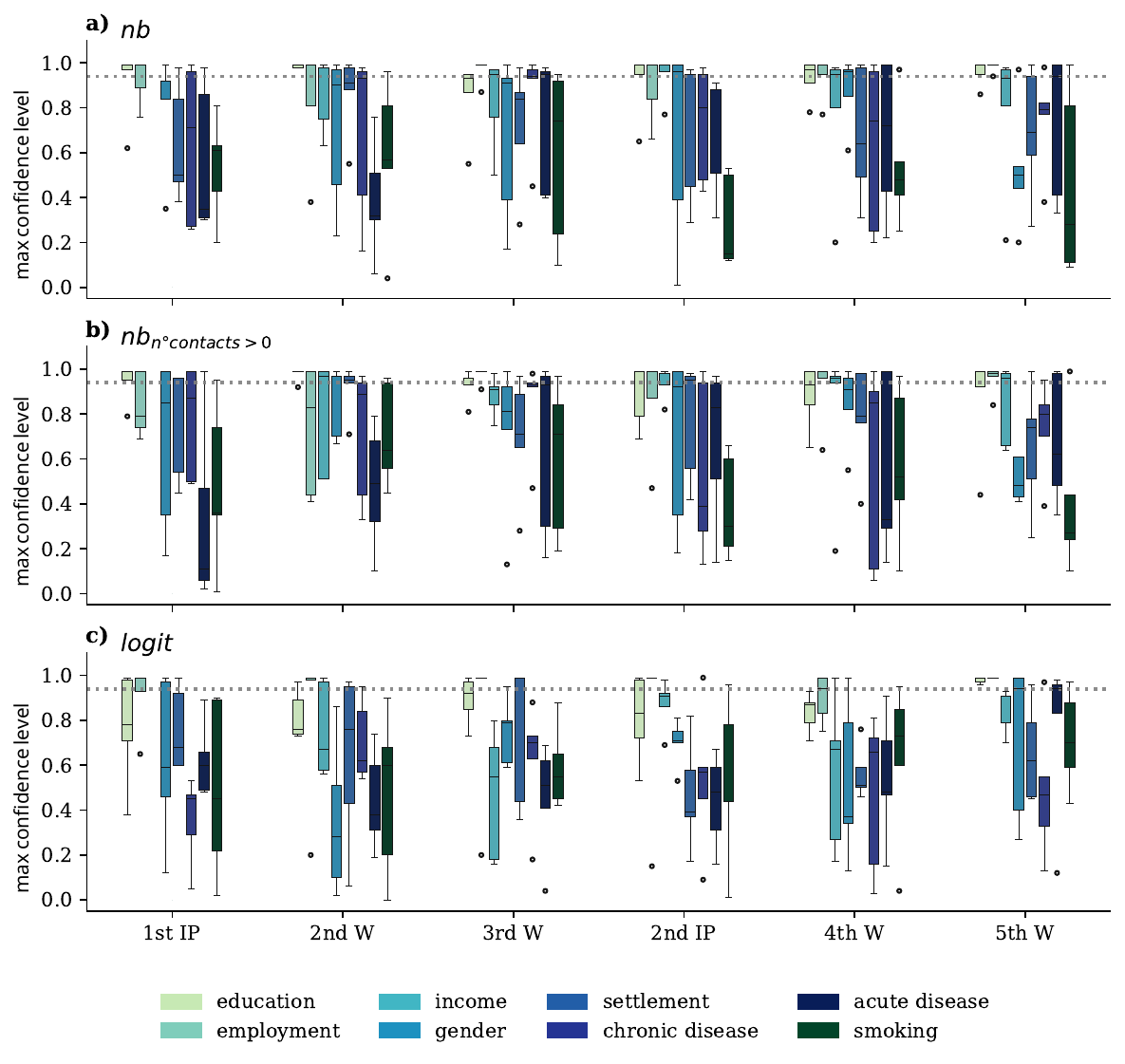}
  \caption{Box-plot of the maximum confidence level at which the effect of the different categories of the variable on the dependent variable of contacts becomes significantly different. The dispersion of the box plot refers to the variation of this value over different age groups. Results are shown for \textit{education level, employment situation, income level, gender, settlement, chronic disease, acute disease, and smoking behaviour}. Results for the three models implemented  $nb$ (negative binomial) $(a)$, $nb_{n contacts>0}$ (negative binomial on non-zeros number of contact) $(b)$, and $logit$ (logistic regression) $(c)$.}
  \label{SI_proxy}
\end{figure}

By applying the same methodology as explained in the Method section we computed the \textit{max significance level} by age for each of the variables and period considered in the analysis for model \ref{stat_model_nbzi} and \ref{stat_model_logitzi}. Fig \ref{SI_proxy} shows the results of the three models that we implemented. Interestingly, we can see that the same qualitative patterns result from both the negative binomial models, that is if we include the observations where the number of contacts is 0 ($nb$), o we discard it ($nb_{n contacts>0}$)  (Fig. \ref{SI_proxy}$a,b$). In particular, although the \textit{maximum confidence level} computed with these models differ in terms of variability, education, employment and income seem to remain the most significant dimensions in terms of explaining differences in contact numbers among subgroups of the population. The logistic regression model ($logit$) (Fig. \ref{SI_proxy}$c$), shows similar results regarding the education and employment situation while it indicates that the other variable analyzed are not actually determining if individuals actually have contacts or not.

\clearpage
\subsubsection*{Community contacts}
\begin{figure}[ht!]
  \centering
  \includegraphics[scale =0.55]{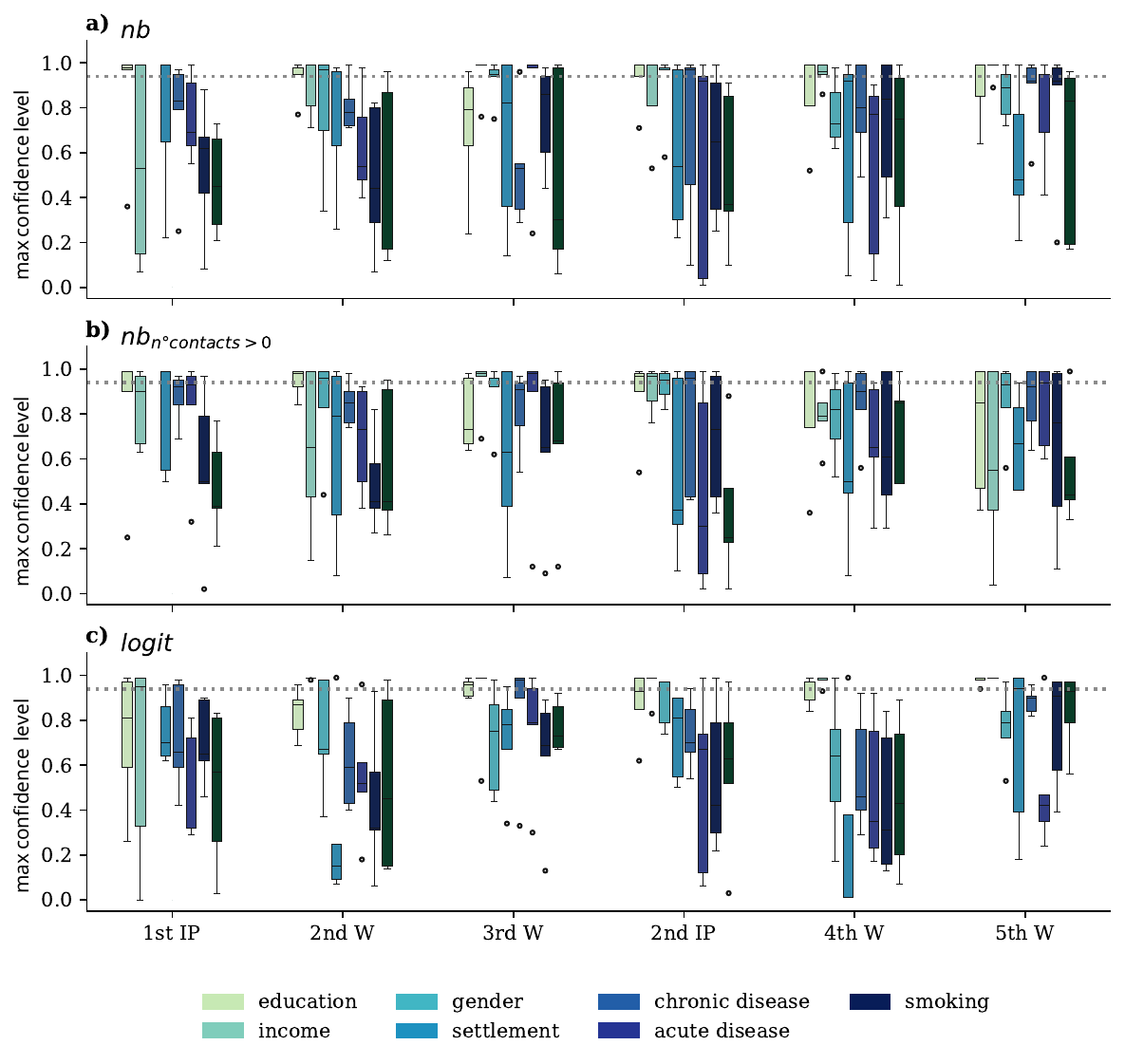}
  \caption{Box-plot of the maximum confidence level at which the effect of the different categories of the variable on the dependent variable of contacts becomes significantly different. The dispersion of the box plot refers to the variation of this value over different age groups. Results are shown for \textit{education level, employment situation, income level, gender, settlement, chronic disease, acute disease, and smoking behaviour}. Results for the three models implemented  $nb$ (negative binomial) $(a)$, $nb_{n contacts>0}$ (negative binomial on non-zeros number of contact) $(b)$, and $logit$ (logistic regression) $(c)$.}
  \label{SI_community}
\end{figure}

Furthermore, we also investigated the contacts happening exclusively in the community layer, excluding the ones happening at work. 
We show here the results of the statistical analysis on the number of contacts in the community layer. Particularly in Fig. \ref{SI_community} we show the results of the three models we are implementing: $1.$ negative binomial regression ($nb$), $2.$ negative binomial regression only on positive observations of the number of contacts  ($nb_{n contacts>0}$), and $3.$ logistic regression to model the probability of having at least one contact ($logit$). In all these models the dependent variable ($y_i$) refers to the community-level contacts of the individuals. Results are similar to the one run considering all the types of contacts together. Indeed, also in this case results indicate that education, employment and income are the most significant dimensions in terms of explaining differences in contact numbers in the community layer among subgroups of the population. Also the logistic regression model ($logit$) (Fig. \ref{SI_community}$c$), shows similar results indicating that only education and employment are significant in determining if individuals actually have contacts or not in the community layer.

\clearpage
\subsection*{Model on vaccination}
We model the probability of getting vaccinated against Covid-19 using a logistic regression model instead of a negative binomial, as the dependent variable was binary.
Namely, we model the probability of getting vaccinated for respondent $i$ using logistic regression as defined in Eq.~\ref{stat_model_vax}:
\begin{equation}
    \log P_i(vax=1) = \alpha + \beta_1 age\_group_i + \beta_2 X_i + \beta_3 age\_group_i* X_i + \epsilon_i
\label{stat_model_vax}
\end{equation}
where $age\_group_i$ is the age class of $i$; $X_i$ is the variable of interest (e.g., education, income  etc.), $age\_group_i*X_i$  is the interaction term of age group and the variable of interest, and $\epsilon_i$ is the error term. 

\begin{figure}[ht!]
  \centering
  \includegraphics[scale =0.6]{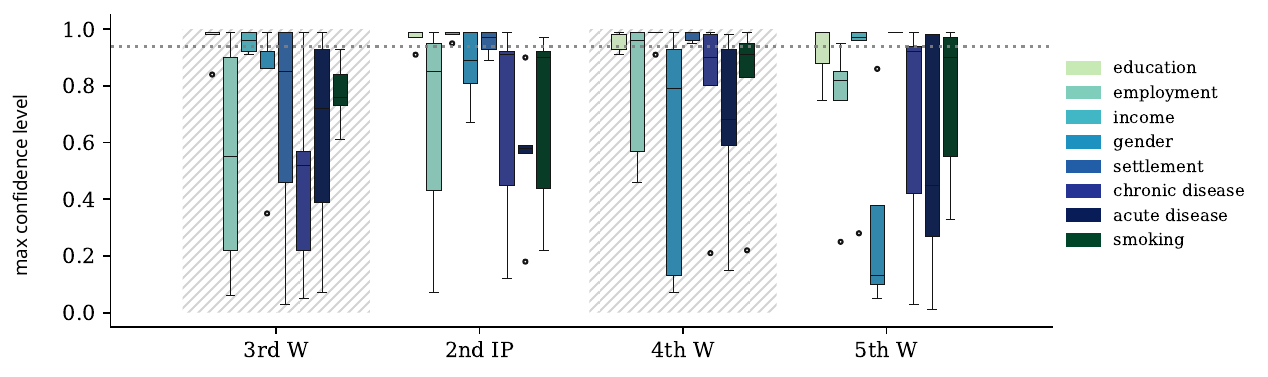}
  \caption{Box-plot of the maximum confidence level at which the effect of the different categories of the variable on vaccination become significantly different. The dispersion of the box plot refers to the variation of this value over different age groups. Results are shown for \textit{education level, employment situation, income level, gender, settlement, chronic disease, acute disease, and smoking behaviour}.}
  \label{SI_stat_vax}
\end{figure}

By applying the same methodology as explained in the Method section we computed the \textit{max significance level} by age for each of the variables and period considered in the analysis (Fig. \ref{SI_stat_vax})

\section*{Contacts}
\subsection*{Average number of contacts in the community and in the work layer by sub-groups}

\begin{figure}[ht!]
  \centering
  \includegraphics[scale =0.4]{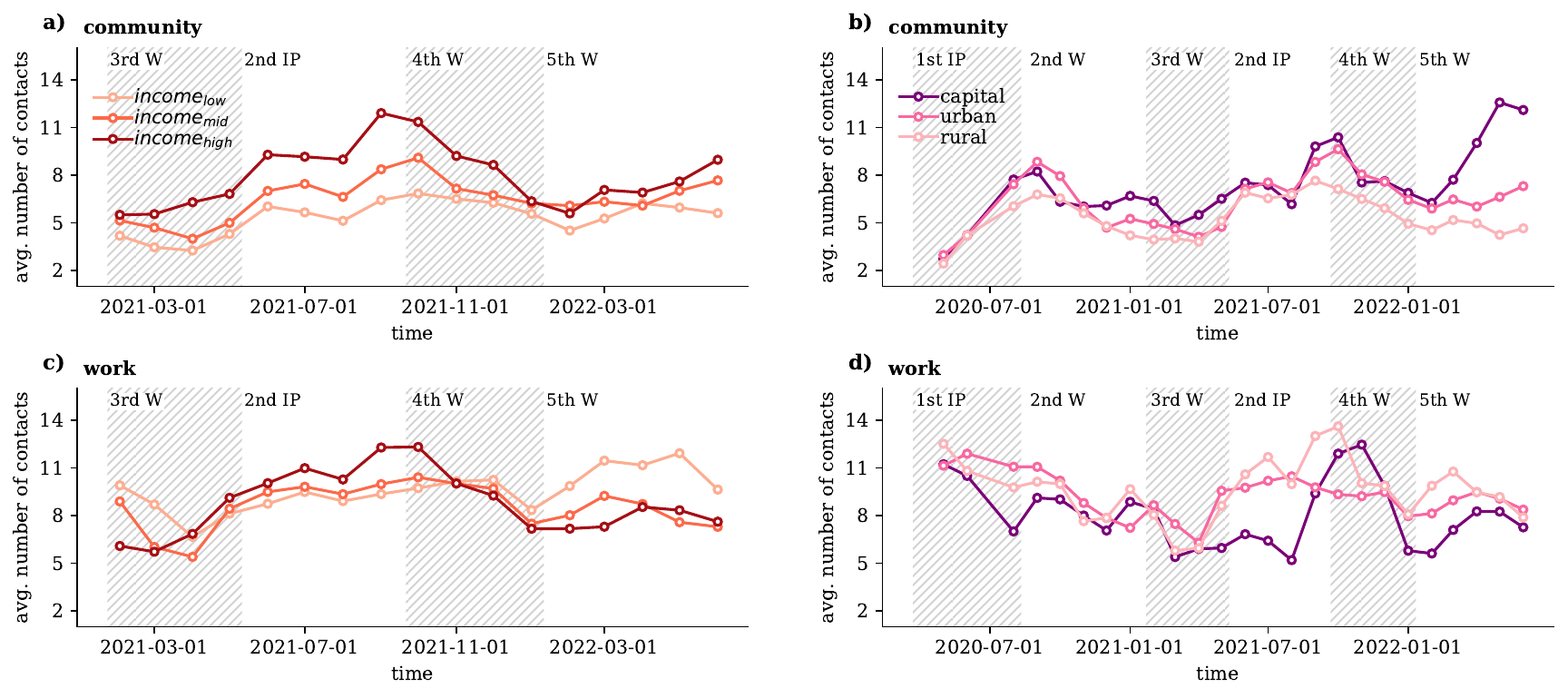}
  \caption{\textbf{(a-b)} Smoothed average number of contacts in the community layer by different income levels $a)$ and settlement types $b)$. \textbf{(c-d)} Smoothed average number of contacts at the workplace by different income levels $c)$ and settlement types $d)$.}
  \label{SI_contacts}
\end{figure}

In the main text, we show the evolution of contacts over time decoupled by education level and employment. For completeness, here we report the same figures for income level Fig \ref{SI_contacts}-a,c and settlement Fig \ref{SI_contacts}-b,d.
Looking at the community contact we can observe that there is a clear rank among the income levels in their number of contacts, with high-income individuals having the highest number of contact and low-income individuals having the lowest. The same is arguable for the individuals living in the capital, which appear to be the most active in the community layer, while, individuals living in rural areas are the less active.
To what concern the contacts at work, we can clearly observe that high-income individuals were the ones who would better adapt to the epidemiological situation, while mid and low income maintained a fairly stable number of contact over time at the workplace. Particularly, they reported a higher number of contacts at the workplace during the Covid-19 waves.  A similar conclusion can be drawn when we decoupled individuals according to their settlement. In this case, individuals living in rural areas appear to be the most active in the workplace while the ones living in the capital tend to have a lower number of interactions in the workplace.  

\subsection*{Average number of contacts in the community layer by sub-groups and age groups}
\begin{figure}[ht!]
  \centering
  \includegraphics[scale =0.35]{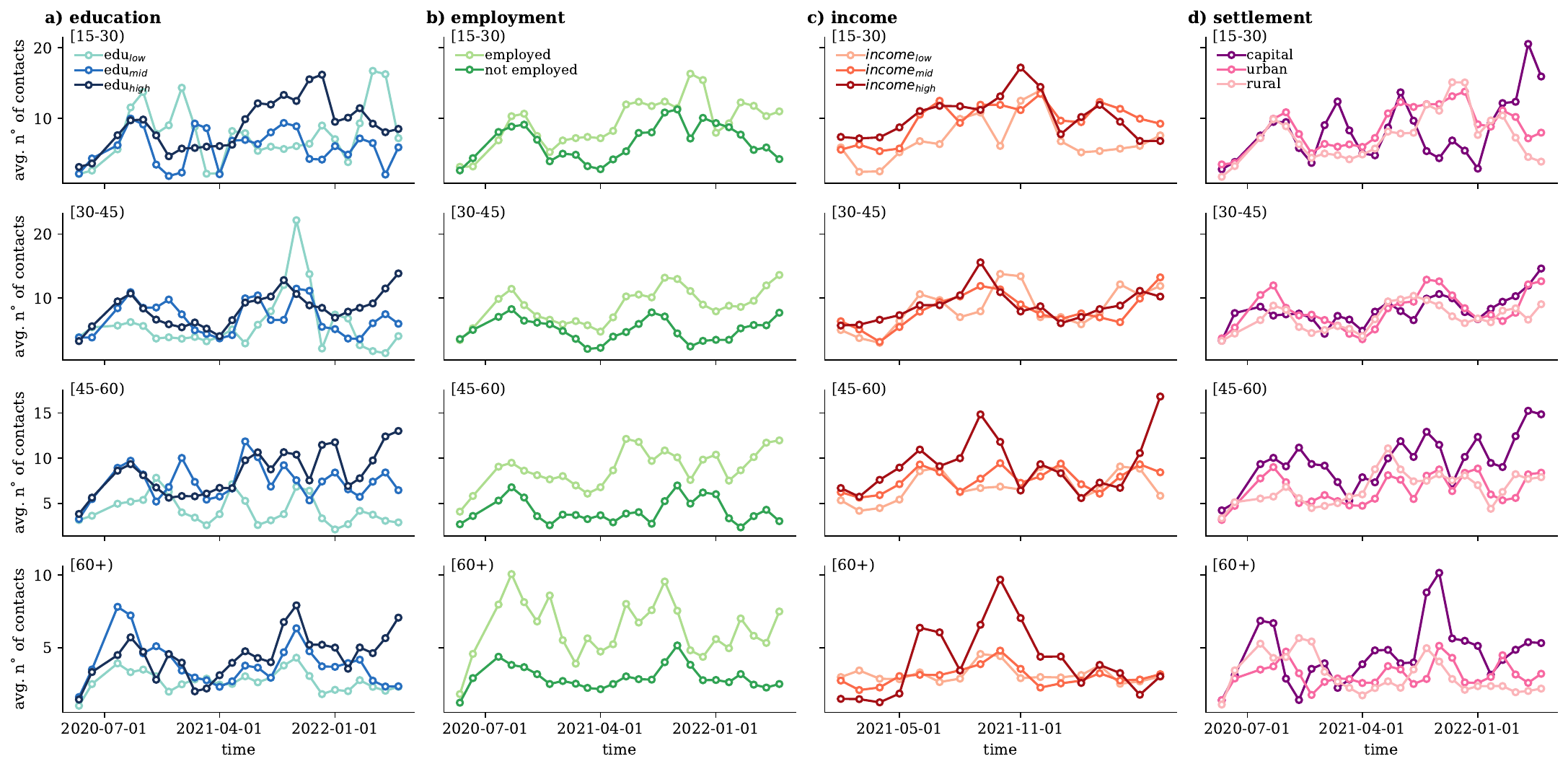}
  \caption{Average number of contacts in the community layer by different age groups\textit{(rows)} and sub-groups \textit{(columns}: education levels $a)$, employment situation $b$, income levels $c)$ and settlement $d)$. }
  \label{SI_contacts_byage}
\end{figure}

To show the robustness of our finding over different age groups in Fig \ref{SI_contacts_byage} we report the evolution of community contacts over time decoupled by education level, employment, income level and settlement. While the correlation with age influences the magnitude of differences among the examined sub-groups, the conclusion discussed in the main text appears to be still valid.

\subsection*{Contact Matrices}
\begin{figure}[ht!]
  \centering
  \includegraphics[scale =0.55]{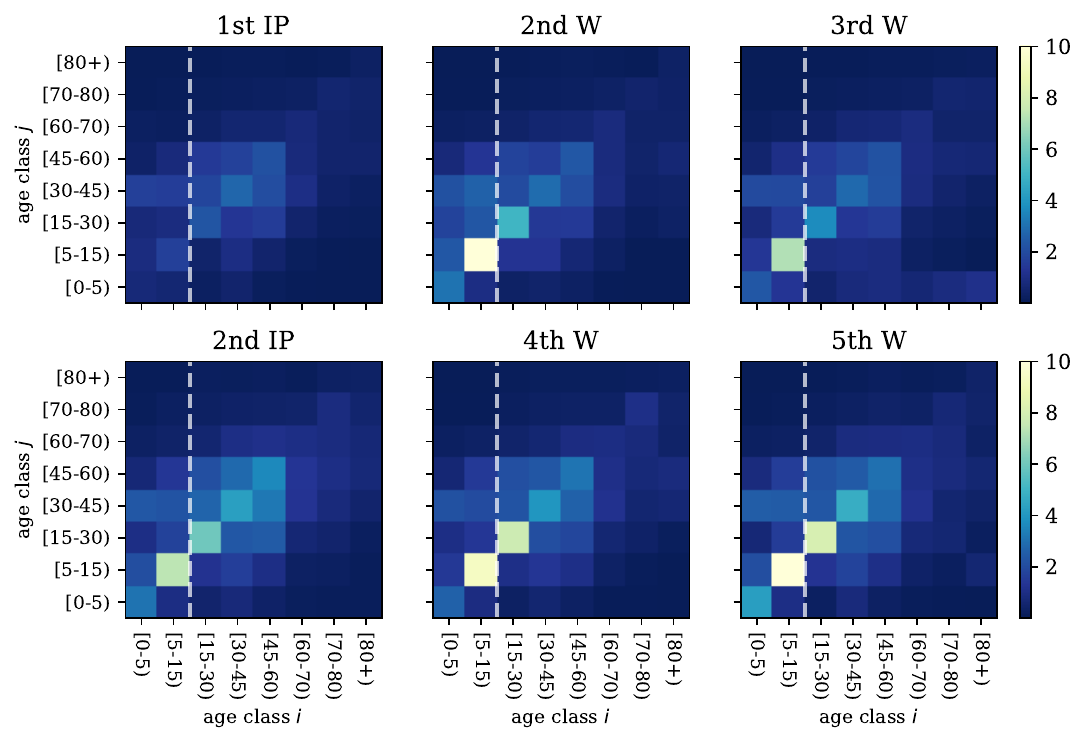}
  \caption{Age contact matrices ($C_{ij}$) for different periods.}
  \label{age_M}
\end{figure}

\begin{figure}[ht!]
  \centering
  \includegraphics[scale = 0.55]{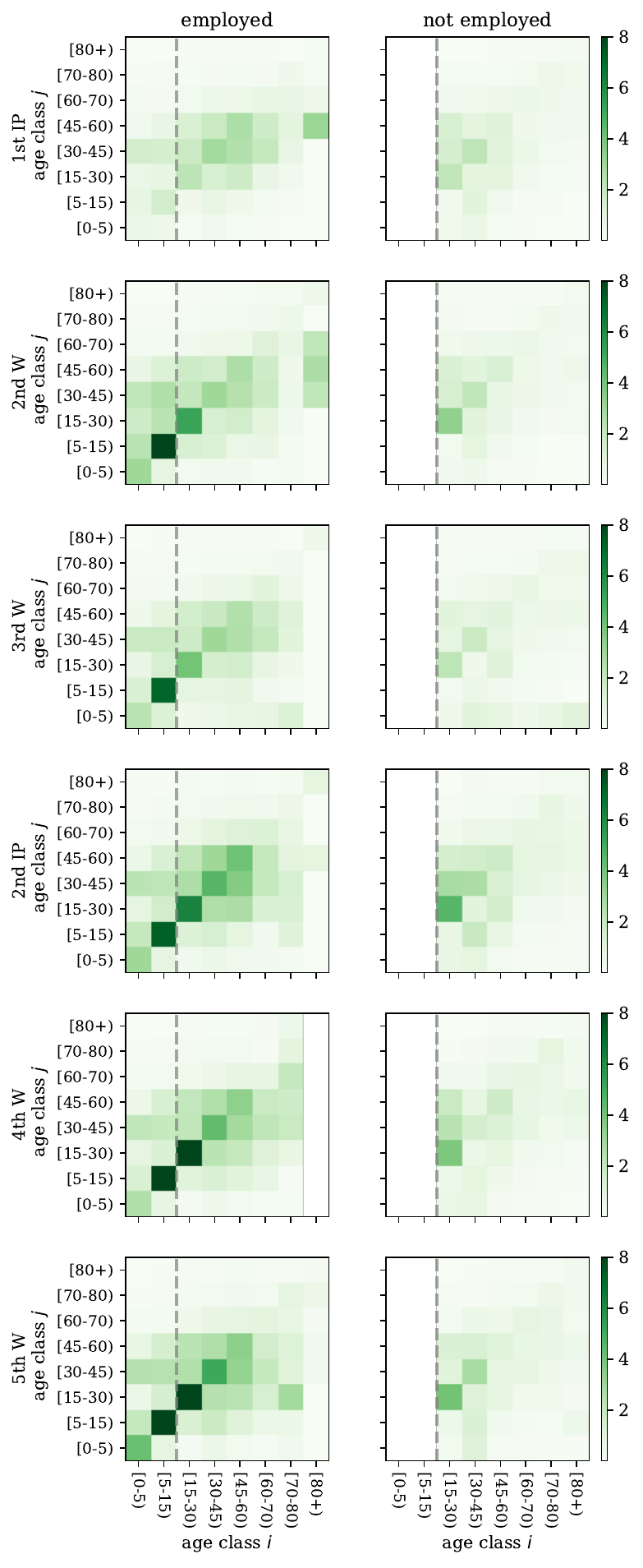}
  \caption{Age contact matrices decoupled by employment situation ($C_{employment i,j}$) for different periods.}
  \label{emp_M}
\end{figure}

\begin{figure}[ht!]
  \centering
  \includegraphics[scale = 0.55]{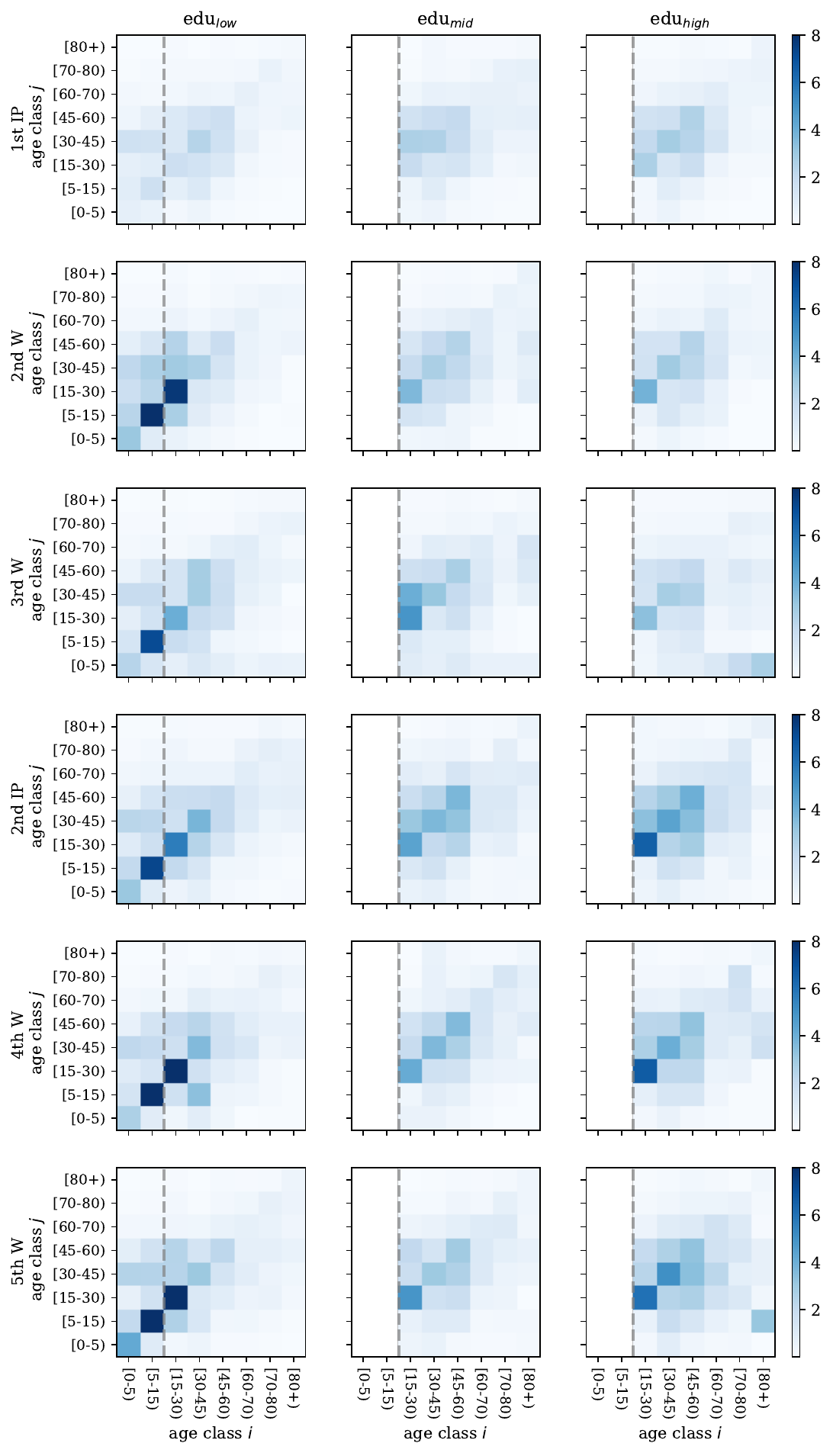}
  \caption{Age contact matrices decoupled by education level ($C_{education i,j}$) for different periods.}
  \label{edu_M}
\end{figure}

\begin{figure}[ht!]
  \centering
  \includegraphics[scale = 0.55]{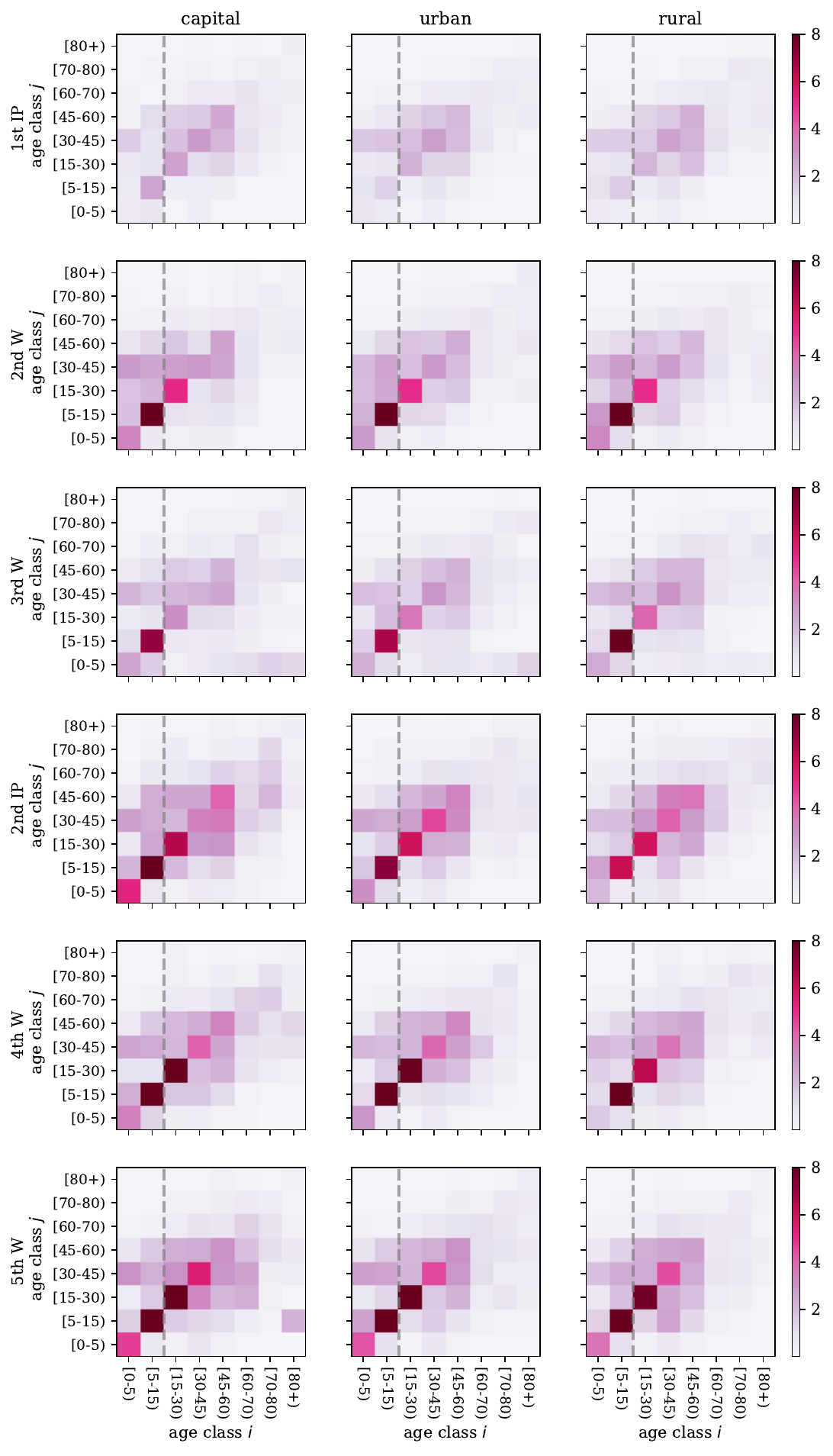}
  \caption{Age contact matrices decoupled by settlement ($C_{settlement i,j}$) for different periods.}
  \label{sett_M}
\end{figure}

\begin{figure}[ht!]
  \centering
  \includegraphics[scale = 0.55]{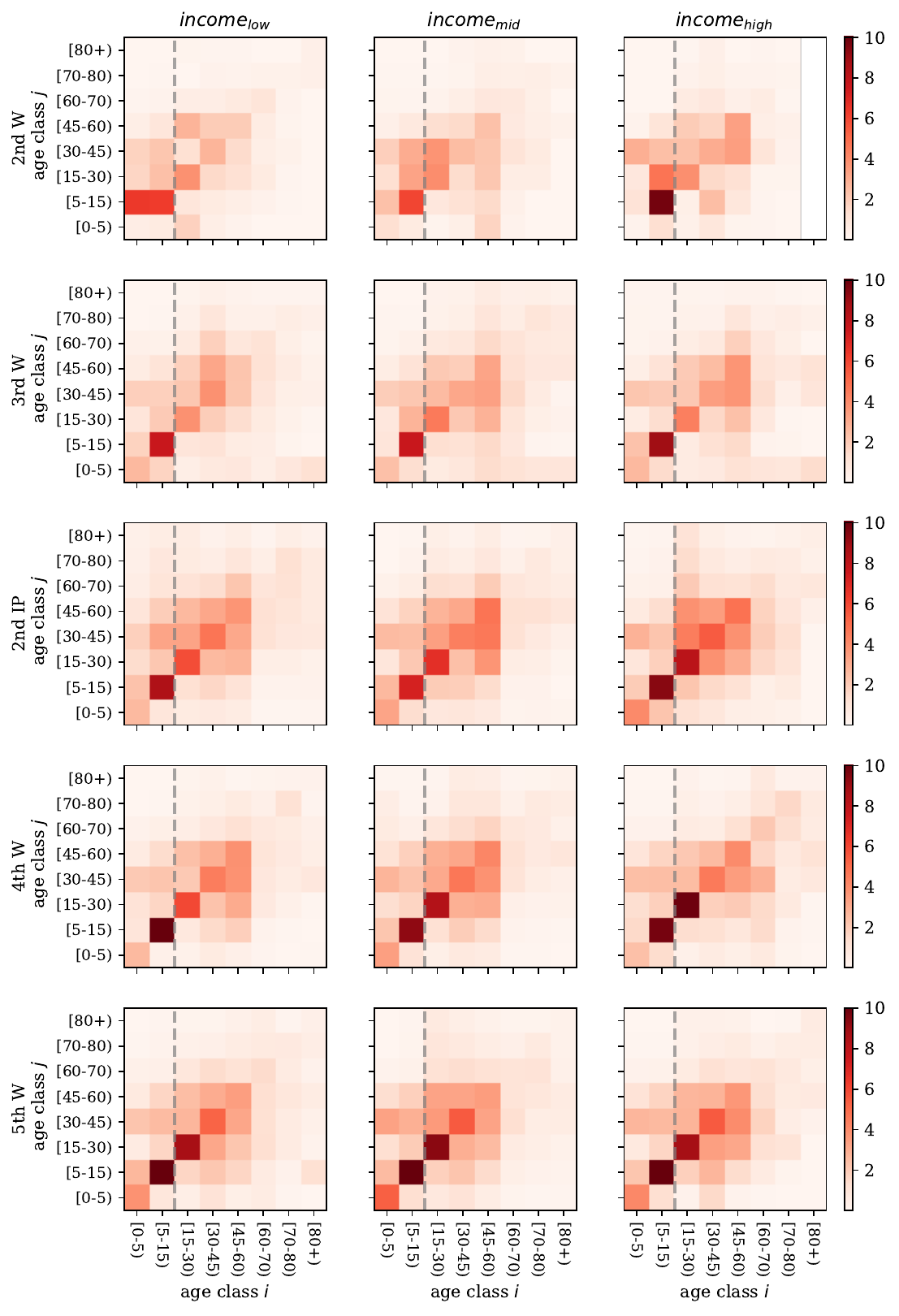}
  \caption{Age contact matrices decoupled by income level ($C_{income i,j}$) for different periods.}
  \label{income_M}
\end{figure}

For each of the periods considered in this study, we computed the age contact matrix $C_{ij}$ considering the whole population as shown in Fig \ref{age_M}. These are the matrices that have been fed to to the \textit{conventional} SEIR model. Instead for the \textit{extended} SEIR model, we computed the decoupled contact matrices ($C_{di,j}$) considering different dimensions as well: $(i)$ employment situation (Fig \ref{emp_M}), $(ii)$ education level (Fig \ref{edu_M}),$(iii)$ settlement (Fig \ref{sett_M}), and $(iv)$ income level (Fig \ref{income_M}).All the matrices have been computed considering the contact at work, in the community and in the household. Detailed explanation of the computation of such matrices is provided in the Method section.

\clearpage
\section*{Vaccination uptake}
\begin{figure}[ht!]
  \centering
  \includegraphics[scale = 0.55]{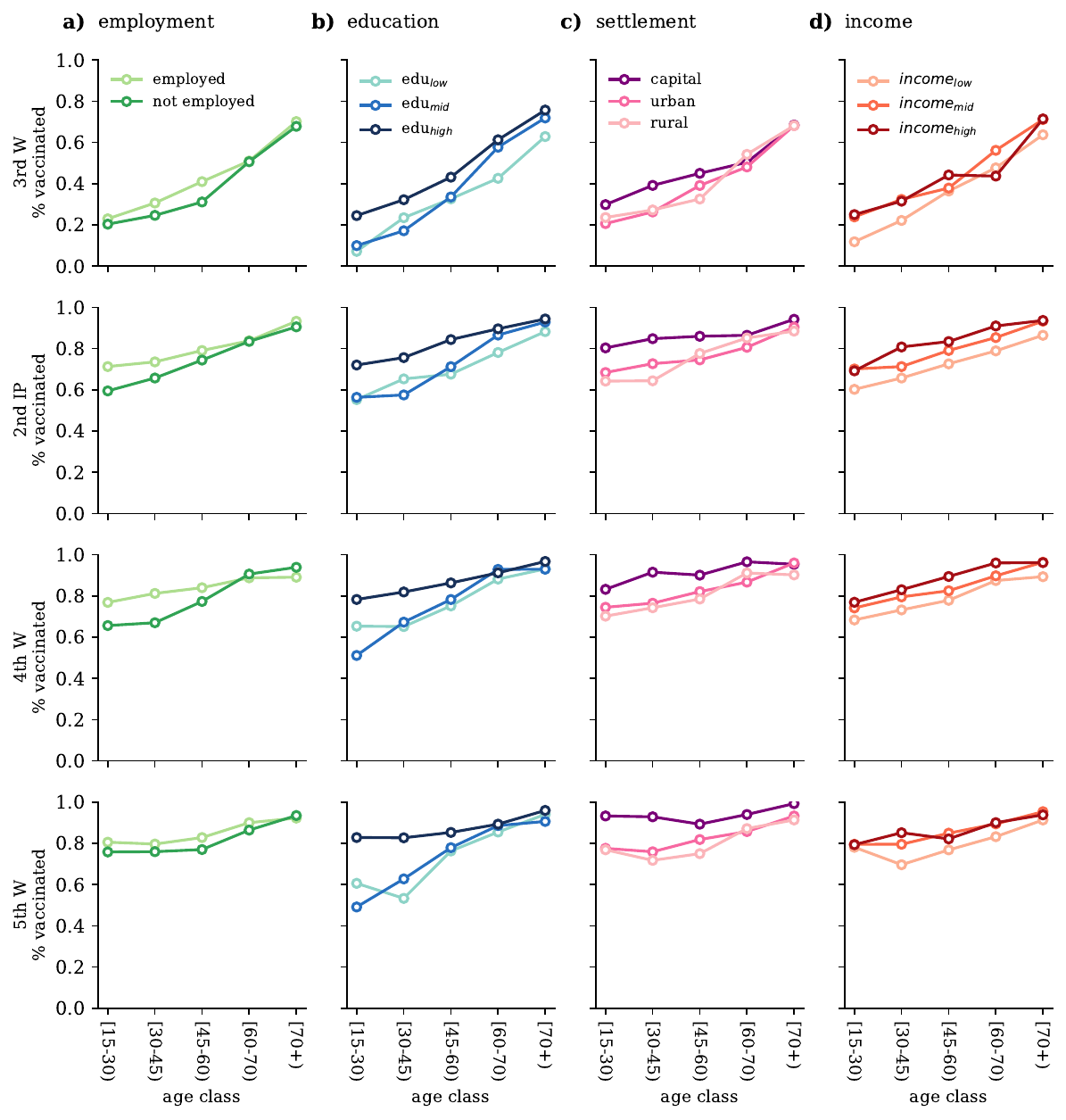}
  \caption{Probability of getting vaccinated given age and $(i)$ employment situation, $(ii)$ education level,$(iii)$ settlement, and $(iv)$ income level}
  \label{SI_vax}
\end{figure}

Here we show the probability of getting vaccinated against COVID-19 given age and another dimension of interest. Namely, we consider $(i)$ employment situation, $(ii)$ education level,$(iii)$ settlement, and $(iv)$ income level. From Fig. \ref{SI_vax} we can observe that privilege groups of the population tend to have higher vaccination uptake across all age groups. This fining is consistent over the four different periods considered in the analysis. 

\section*{Epidemic models}
In this section, we report the ODE's equation of the $(i)$ \textit{conventional} age-stratified SEIRD model and, $(i)$ \textit{extended} SEIRD where beyond the age stratification we can differentiate the population along others dimension of interest $(\bar{d})$. Specifically, the reported equations for the \textit{extended} SEIRD model account also for vaccination.

\subsection*{\textit{Conventional} age-stratified SEIRD}
\label{sec_classical_SEIRD}
Let's  consider an infectious  disease that can be described with a Susceptible-Exposed-Infected-Recovered-Death model \cite{rohani}. The epidemic dynamic is encoded in the set of differential equations in Eq. \ref{SEIRD_class}. Where, $i$ indicates the age group of the ego, $j$ indicates the age group of the alter,  $\beta$ is the probability of transmission given a contact, $\epsilon$ is the rate at which individuals become infectious, $\mu$ is the recovery rate, $C_{ij}$ is the age contact matrix, and $IFR_i$ is the infection fatality rate by age group.

\begin{equation}
\label{SEIRD_class}
\begin{array}{lcl} 
\dot{S}_i &=& -\lambda_i S_i \\
\dot{E}_i &=& \lambda S_i -\epsilon E_i \\
\dot{I}_i &=& \epsilon E_i-\mu I_i \\
\dot{R}_i &=& \mu I_i \\
\dot{D}_i &=& \mu IFR_i I_i
\end{array}
\end{equation}
The force of infection is defined as  in eq. \ref{foi_classical}

\begin{equation}
\lambda_{i}=  \beta \sum_{j} \frac{C_{ij}}{N_{j}} I_{j}
\label{foi_classical}
\end{equation}

\subsection*{\textit{Extended SEIRD} with vaccination}

We extend the \textit{conventional SEIRD} model to account for different vaccination uptake of different groups of the population. Namely, each of the compartments is now considered separately for vaccinated and unvaccinated individuals. We define as $g_1$ and $g_2$ the efficiency of the vaccination respectively against  infection and against death. In addition, we consider the delay in the official registrations of deaths by adding a new compartment $Da$ and a delay of $\Delta^{-1}$ days. The equations of the model are presented in equation \ref{SEIRD_v_ad}.

\begin{equation}
\label{SEIRD_v_ad}
\begin{array}{lcl} 
\dot{S}_{\bar{d},i} & = & -\lambda_{\bar{d},i} S_{\bar{d},i} \\ 
\dot{Sv}_{\bar{d},i} & = & -(1-g_1)\lambda_{\bar{d},i} Sv_{\bar{d},i}\\ 

\dot{E}_{\bar{d},i} & = & \lambda_{\bar{d},i} S_{\bar{d},i} -\epsilon E_{\bar{d},i} \\ 
\dot{Ev}_{\bar{d},i} & = & (1-g_1)\lambda_{\bar{d},i} Sv_{\bar{d},i}-\epsilon Ev_{\bar{d},i} \\

\dot{I}_{\bar{d},i} & = & \epsilon E_{\bar{d},i} - \mu I_{\bar{d},i} \\
\dot{Iv}_{\bar{d},i} & =& \epsilon Ev_{\bar{d},i}- \mu Iv_{\bar{d},i} \\

\dot{R}_{\bar{d},i} & = &  \mu (1-IFR_i) I_{\bar{d},i} \\ 
\dot{Rv}_{\bar{d},i} & =&  \mu (1-(1-g_2)IFR_i) Iv_{\bar{d},i} \\

\dot{D}_{\bar{d},i} & = &  \mu IFR_iI_{\bar{d},i} \\ 
\dot{Dv}_{\bar{d},i} & =&   \mu (1-g_2) IFR_i Iv_{\bar{d},i}\\

\dot{D_a}_{\bar{d},i} & = &  \Delta^{-1}\dot{D}_{\bar{d},i} \\ 
\dot{Dv_a}_{\bar{d},i} & =&  \Delta^{-1}\dot{Dv}_{\bar{d},i}\\

\end{array}
\end{equation}

The force of infection is defined as  in eq. \ref{foi_vacc}

\begin{equation}
\begin{array}{lcl} 
\lambda_{\bar{d},i}(t)=  \beta \sum_{j} \frac{C_{\bar{d},ij}}{N_{j}} [{I_{j}}+Iv_{j}]
\end{array}
\label{foi_vacc}
\end{equation}

\section*{Epidemic Simulations}

We developed stochastic, discrete-time, compartmental models using chain binomial processes to simulate the transitions among compartments. Specifically, at each time step $t$, the model samples the number of individuals in group $(\bar{d},i)$ and compartment $X$ transitioning to compartment $Y$ from $PrBin(X_{\bar{d},i}(t),p_{X_{\bar{d},i}\xrightarrow{
}Y_{\bar{d},i}}(t))$. Here, $p_{X_{\bar{d},i}\xrightarrow{}Y_{\bar{d},i}}(t)$ represents the transition probability.

To illustrate this, let's consider the number of individuals in the group $(\bar{d}, i)$ and compartment $S$ that at time $t$ become exposed transiting to compartment $E$. Thus, the number of individuals in $S_{\bar{d}, i}(t)$ getting exposed are extracted from a $PrBin(S_{\bar{d}, i}(t), \lambda_{\bar{d}, i}(t))$ where $ \lambda_{\bar{d}, i}(t)$ is the \textit{force of infection}.

The model has been initialize by computing the population distributions from the MASZK data, while we set the Hungarian population size to 9.750.000.
The decoupled contact matrices have been computed considering the contacts happening at work, in the community and in with family members. 
While, the epidemiological parameters are set to realistic values to closely simulate the characteristics of Covid-19. These values are retrieved from the literature. In particular, $\epsilon$ is set to $2.4$; $\gamma$ is set to $1/6.6$ \cite{gozzi2022anatomy, kissler2020projecting, backer2020incubation}.
The transmission rate $\beta$ is computed in each of the periods using the Next Generation Matrix approach \cite{blackwood2018introduction} on the aggregate age-contact matrices corresponding to the periods analysed. We fixed $R0 = 2.5$ and we derived $\beta$ using Eq. \ref{NGM}.

\begin{equation}
\label{NGM}
    R_0 = \frac{\beta}{\mu} \rho(C_{ij})
\end{equation}

Where $\rho(C_{ij})$ is the spectral radius of the mage contact matrix. 
In the simulations in which we introduce vaccination, we respectively set the $g_1 =0.6$ and $g_2=0.8$ \cite{voko2022nationwide,shapiro2021efficacy}.
The initial size of the epidemic is set to 5. All the results in the main text refer to the median over 500 simulations of the model.

\subsection*{Impact of different contact patterns}
\begin{figure}[ht!]
  \centering
  \includegraphics[scale = 0.41]{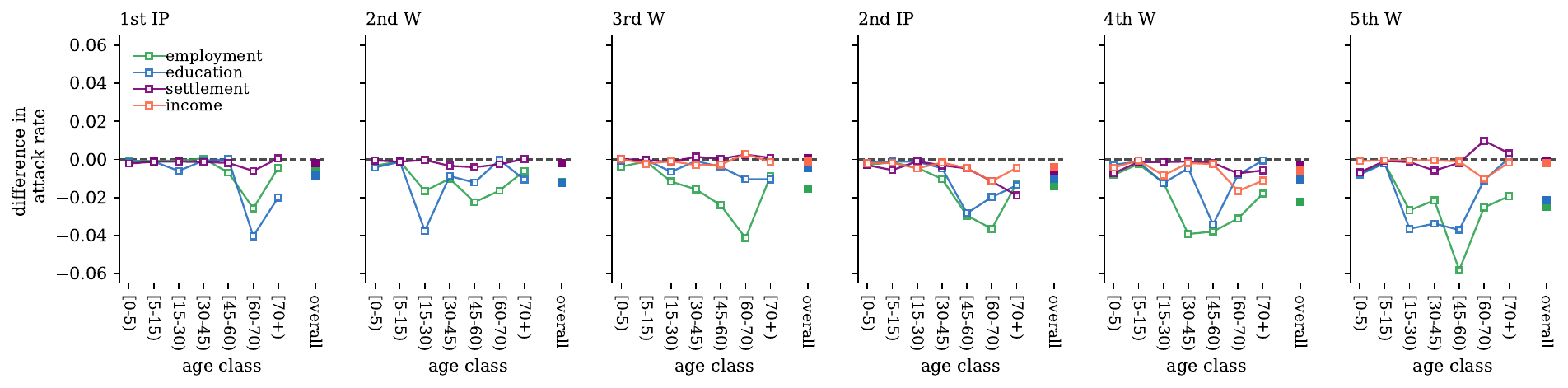}
  \caption{Averted attack rate by age and $(i)$ employment situation, $(ii)$ education level, $(iii)$ settlement, $(iv)$ income and $(v)$ overall, in the different periods (\textit{columns})}.
  \label{SI_ARdiff}
\end{figure}

In this section, we show the results of the \textit{extended SEIR} model when differences in age contact matrices are considered for different sub-groups of the population. Particularly, we model age contact interactions differentiating individuals along their employment situation, education level, settlement and income level. For each of the dimensions considered, we run the \textit{extended} SEIR model. We look at $(i)$ and how the prediction of this model differs from the \textit{conventional SEIR} and $(ii)$ how the attack rate differs for different subgroups as a result of their differences in contact patterns.
Specifically, in Fig. \ref{SI_ARdiff} we show the difference between attack rate by age group as predicted by the \textit{extended SEIR} model and the \textit{conventional SEIR} model. The results are shown for each of the periods considered in the analysis. Again, as demonstrated in the main text, the \textit{conventional SEIR} model tend to overestimate the attack rate by age group with respect to \textit{extended SEIR} model, particularly when employment situation and education are taken into account.

\begin{figure}[ht!]
  \centering
  \includegraphics[scale = 0.55]{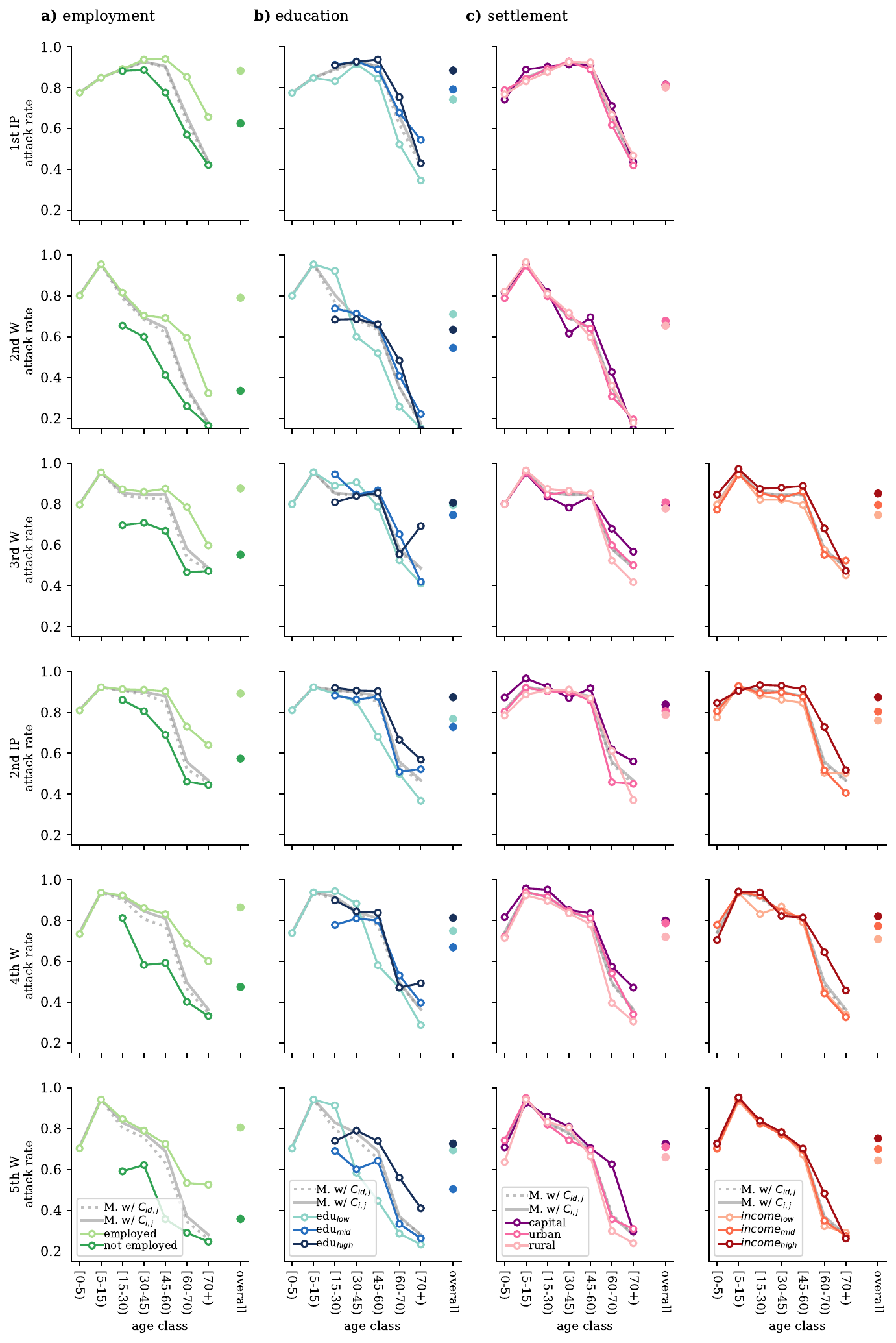}
  \caption{Attack rate by age and employment situation, education level, settlement and income (\textit{columns}), in different periods (\textit{rows})}
  \label{SI_AR_modelC_idj}
\end{figure}

In Fig. \ref{SI_AR_modelC_idj} we show the output of each of the \textit{extended SEIR} models in terms of attack rate by age by differentiating along the subgroups taken into account. Results are shown for each of the periods considered in the analysis. 

As shown in the main text the analysis over the other periods confirms that, employed and highly educated individuals happened to be the most infected groups in all age groups. When decoupling age contact matrices by settlement and income, although differences appear smaller between groups, high-income individuals and the ones living in the capital are more infected, particularly elderly ones with age 60+.

\clearpage
\subsection*{Impact of different vaccination uptake}
\begin{figure}[ht!]
  \centering
  \includegraphics[scale = 0.5]{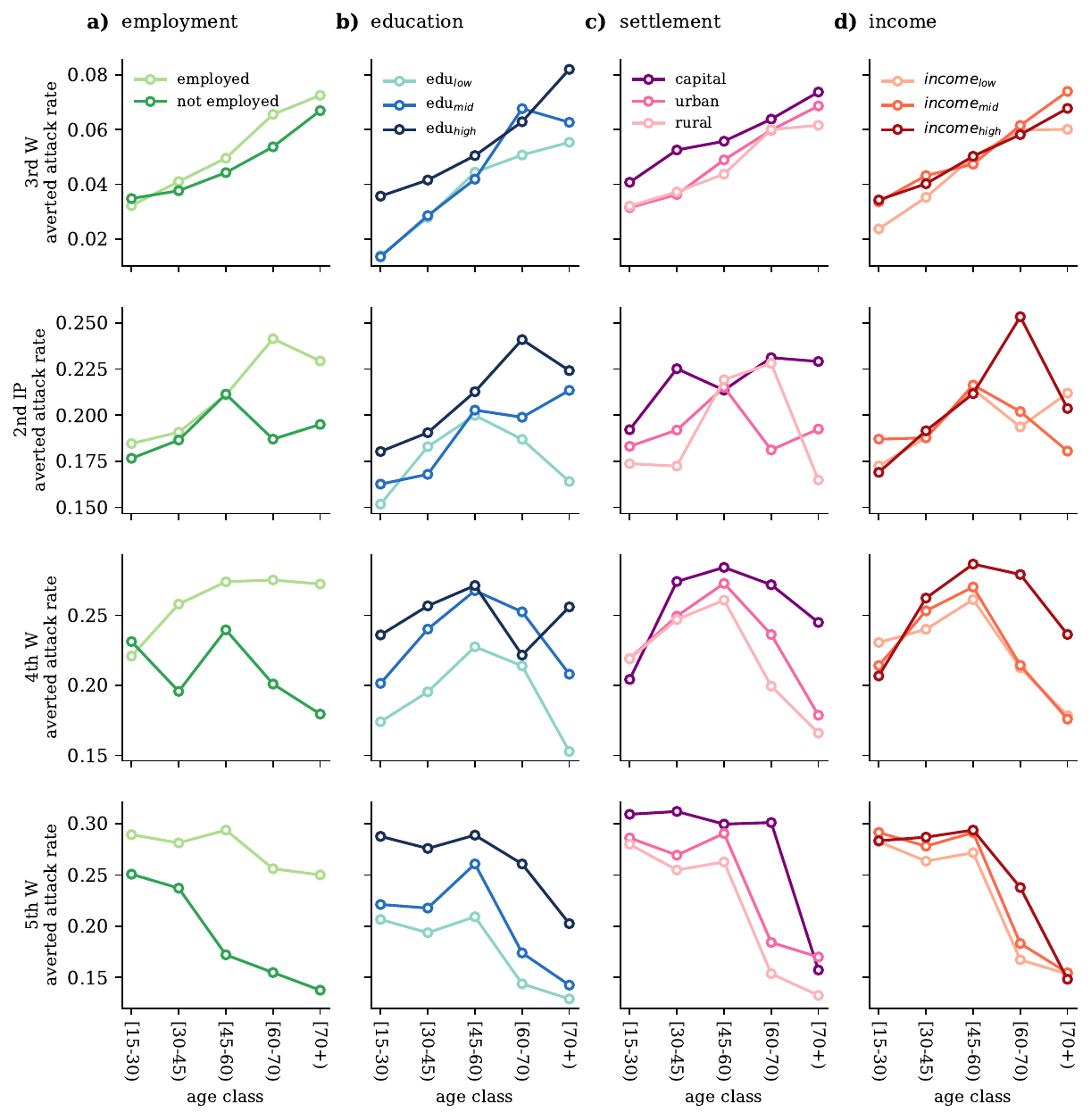}
  \caption{Difference in attack rate due to vaccination by age and employment situation, education level, settlement and income (\textit{columns}), in different periods (\textit{rows})}
  \label{SI_vax_diffAR}
\end{figure}

In order to show the impact of different vaccination uptake here we show the \textit{averted attack rate} by age  due to vaccination (Fig. \ref{SI_vax_diffAR}). Specifically, we show the difference among the attack rate by age, for the different subgroups as predicted by the \textit{extended SEIR} in the non-vaccination scenario with respect to the one in which individuals get vaccinated according to their age and subgroup- as shown in Fig. \ref{SI_vax}. 

The findings from the additional periods support the observations discussed in the main text.  Indeed, Fig. \ref{SI_vax_diffAR} clearly shows that vaccination benefits are disproportionately advantageous for more privileged population groups.

\section*{Model Calibration}
We calibrate a SEIRD model with vaccination by modelling differences in contacts patterns and vaccination uptake among employed and not employed individuals belonging to different education levels. 
In particular, we calibrate the free parameters of the model using an Approximate Bayesian Computation (ABC) technique \cite{minter2019approximate,sunnaaker2013approximate} . First, we define the prior distributions of the free parameters $P(\theta)$, a number of accepted sets $N$, an error metric $m(E,E')$, and a tolerance $\delta$. We start sampling a set of parameters $\theta$ from $P(\theta)$, and generate an instance of the model using these parameters. Then, using the chosen error metric we compare an output quantity $E'$ of the model with the corresponding real quantity $E$: if $m(E,E')$ $< \delta$ then we accept the set $\theta$, otherwise we reject it. We repeat this accept/reject step until $N$ parameter sets are accepted. The empirical distribution of the accepted sets is an approximation of their real posterior distribution. Finally, we generate an ensemble of possible epidemic trajectories sampling parameter sets from the posteriors distributions. In this work, we consider the following free parameters and prior distributions:

\begin{itemize}
    \item the transmission rate parameter $\beta$: the prior distribution is set to $U(0.02,0.15)$
    \item the delay in reporting deaths $\Delta$: the prior distribution is set to $U(5,20)$
    \item the initial recovered population $R$: the prior distribution is set to $U(700K,2800K)$  
    \item the initial exposed population $E$: the prior distribution is set to $U(100,3K)$ 
    \item the initial infected population   $I$: the prior distribution is set to $U(200,9K)$
\end{itemize}

\begin{figure}[ht!]
  \centering
  \includegraphics[scale = 0.45]{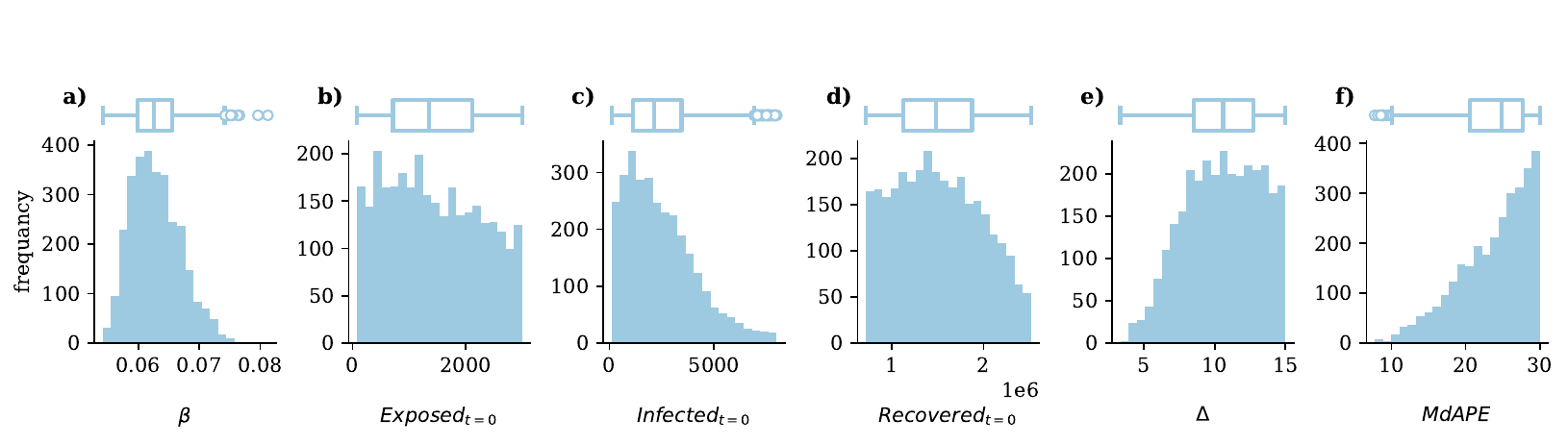}
  \caption{\textbf{(a-e)}Posterior distribution of the calibrated parameters of the model \textbf{(f)} Distribution of the MdAPE ($\epsilon$). }
  \label{SI_params}
\end{figure}

We calibrate our model on the aggregate number of daily deaths from 09/2021 to 01/2022.
For simplicity, as the percentage of those who were vaccinated was quite stable \cite{statista} in the period considered we assume that the population got vaccinated at time $0$.
As an error metric, we use the \textit{Median Absolute Percentage Error (MdAPE)}. We also set the number of accepted sets $N = 3000$ and the tolerance $\delta = 0.3$  In Fig \ref{SI_params} are shown the posterior distributions of the parameters calibrated through the ABC rejection algorithm. 

The fixed parameters of the model have been informed from the literature. In particular:

\begin{itemize}
    \item the efficacy of the vaccine against infection and against death, is modelled as a normal distribution with mean respectively $g_1 = 0.7$, $g_2 = 0.8$, and standard deviation $0.05$ \cite{voko2022nationwide,shapiro2021efficacy}. This choice has been made to account for the variability of the efficacy of the different vaccination types and against the different variants. 
    \item the recovery rate $\mu = 1/2.5$ \cite{gozzi2022anatomy,backer2020incubation,kissler2020projecting}
    \item the incubation period $\epsilon = 1/4$ \cite{gozzi2022anatomy,backer2020incubation,kissler2020projecting}
    \item the infection fatality rate by age $IFR_i$ is set as in Table \ref{ifr_age} \cite{salje2020estimating}
\end{itemize}

\begin{table}[h!]
\centering
\begin{tabular}{@{}lllllllllllllll@{}}
\cmidrule(lr){6-10}
 &  &  &  &  &  & \textbf{Age group} &  & \textbf{IFR}     &  &  &  &  &  &  \\ \cmidrule(lr){6-10}
 &  &  &  &  &  & {[}0-5)   &  & 0.001\% &  &  &  &  &  &  \\
 &  &  &  &  &  & {[}5-15)  &  & 0.001\% &  &  &  &  &  &  \\
 &  &  &  &  &  & {[}15-30) &  & 0.005\% &  &  &  &  &  &  \\
 &  &  &  &  &  & {[}30-45) &  & 0.02\%  &  &  &  &  &  &  \\
 &  &  &  &  &  & {[}45-60) &  & 0.2\%   &  &  &  &  &  &  \\
 &  &  &  &  &  & {[}60-70) &  & 0.7\%   &  &  &  &  &  &  \\
 &  &  &  &  &  & {[}70-80) &  & 1.9\%   &  &  &  &  &  &  \\
 &  &  &  &  &  & {[}80+)   &  & 8.3\%   &  &  &  &  &  &  \\ \cmidrule(lr){6-10}
\end{tabular}
\caption{Infection Fatality Rate by age group}
\label{ifr_age}
\end{table}
In Fig. \ref{SI_params} we report the number of daily real and simulated deaths (median and 95\% CI).

\begin{figure}[ht!]
  \centering
  \includegraphics[scale = 0.45]{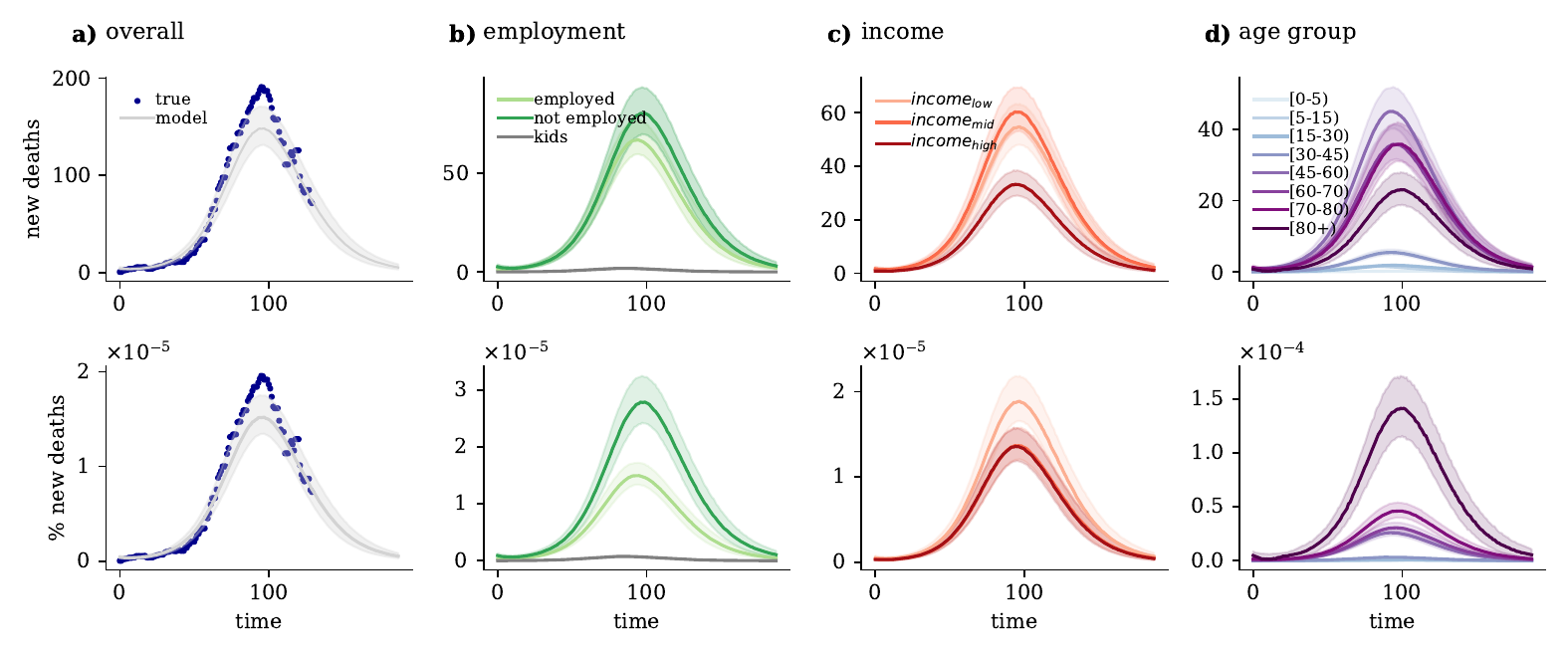}
  \caption{\textbf{(a-d)} Number of daily deaths real (blue dots) and simulated (grey) overall (a), by education (b), employment (c), and age group (d).\textbf{(e-h)} Percentage of daily deaths real(red line) and simulated (grey) overall (e), by education (f), employment (g), and age group(h). Results refers to the median of 3000 simulations. }
  \label{SI_calib_res}
\end{figure}


\begin{thebibliography}{10}

\bibitem{marmot2008closing}
Michael Marmot, Sharon Friel, Ruth Bell, Tanja~AJ Houweling, and Sebastian
  Taylor.
\newblock Closing the gap in a generation: health equity through action on the
  social determinants of health.
\newblock {\em The lancet}, 372(9650):1661--1669, 2008.

\bibitem{mamelund2021association}
Svenn-Erik Mamelund, Clare Shelley-Egan, and Ole Rogeberg.
\newblock The association between socioeconomic status and pandemic influenza:
  systematic review and meta-analysis.
\newblock {\em PLoS One}, 16(9):e0244346, 2021.

\bibitem{kikuti2015spatial}
Mariana Kikuti, Geraldo~M Cunha, Igor~AD Paploski, Amelia~M Kasper, Monaise~MO
  Silva, Aline~S Tavares, Jaqueline~S Cruz, T{\'a}ssia~L Queiroz, Moreno~S
  Rodrigues, Perla~M Santana, et~al.
\newblock Spatial distribution of dengue in a brazilian urban slum setting:
  role of socioeconomic gradient in disease risk.
\newblock {\em PLoS neglected tropical diseases}, 9(7):e0003937, 2015.

\bibitem{mena2021}
Gonzalo~E. Mena, Pamela~P. Martinez, Ayesha~S. Mahmud, Pablo~A. Marquet,
  Caroline~O. Buckee, and Mauricio Santillana.
\newblock Socioeconomic status determines covid-19 incidence and related
  mortality in santiago, chile.
\newblock {\em Science}, 372(6545):eabg5298, 2021.

\bibitem{Burstrom2020}
Bo~Burström and Wenjing Tao.
\newblock {Social determinants of health and inequalities in COVID-19}.
\newblock {\em European Journal of Public Health}, 30(4):617--618, 07 2020.

\bibitem{paul2021socio}
Ayan Paul, Philipp Englert, and Melinda Varga.
\newblock Socio-economic disparities and covid-19 in the usa.
\newblock {\em Journal of Physics: Complexity}, 2(3):035017, 2021.

\bibitem{zhao_harris_ellis_pebody_2015}
H.~Zhao, R.~J. Harris, J.~Ellis, and R.~G. Pebody.
\newblock Ethnicity, deprivation and mortality due to 2009 pandemic influenza
  a(h1n1) in england during the 2009/2010 pandemic and the first post-pandemic
  season.
\newblock {\em Epidemiology and Infection}, 143(16):3375–3383, 2015.

\bibitem{Gozzi2020}
Nicol{\`o} Gozzi, Michele Tizzoni, Matteo Chinazzi, Leo Ferres, Alessandro
  Vespignani, and Nicola Perra.
\newblock Estimating the effect of social inequalities in the mitigation of
  covid-19 across communities in santiago de chile.
\newblock {\em medRxiv}, 2020.

\bibitem{Jay2020}
Jonathan Jay, Jacob Bor, Elaine~O. Nsoesie, Sarah~K. Lipson, David~K. Jones,
  Sandro Galea, and Julia Raifman.
\newblock Neighbourhood income and physical distancing during the covid-19
  pandemic in the united states.
\newblock {\em Nature Human Behaviour}, 12, 12 2020.

\bibitem{valdano2021highlighting}
Eugenio Valdano, Jonggul Lee, Shweta Bansal, Stefania Rubrichi, and Vittoria
  Colizza.
\newblock Highlighting socio-economic constraints on mobility reductions during
  covid-19 restrictions in france can inform effective and equitable pandemic
  response.
\newblock {\em Journal of travel medicine}, 28(4):taab045, 2021.

\bibitem{pullano2020evaluating}
Giulia Pullano, Eugenio Valdano, Nicola Scarpa, Stefania Rubrichi, and Vittoria
  Colizza.
\newblock Evaluating the effect of demographic factors, socioeconomic factors,
  and risk aversion on mobility during the covid-19 epidemic in france under
  lockdown: a population-based study.
\newblock {\em The Lancet Digital Health}, 2(12):e638--e649, 2020.

\bibitem{bonaccorsi2020economic}
Giovanni Bonaccorsi, Francesco Pierri, Matteo Cinelli, Andrea Flori, Alessandro
  Galeazzi, Francesco Porcelli, Ana~Lucia Schmidt, Carlo~Michele Valensise,
  Antonio Scala, Walter Quattrociocchi, et~al.
\newblock Economic and social consequences of human mobility restrictions under
  covid-19.
\newblock {\em Proceedings of the National Academy of Sciences},
  117(27):15530--15535, 2020.

\bibitem{Sommer2015}
Isolde Sommer, Ursula Griebler, Peter Mahlknecht, Kylie Thaler, Kathryn
  Bouskill, Gerald Gartlehner, and Shanti Mendis.
\newblock Socioeconomic inequalities in non-communicable diseases and their
  risk factors: an overview of systematic reviews.
\newblock {\em BMC Public Health}, 15, 09 2015.

\bibitem{Bambra964}
Clare Bambra, Ryan Riordan, John Ford, and Fiona Matthews.
\newblock The covid-19 pandemic and health inequalities.
\newblock {\em Journal of Epidemiology \& Community Health}, 74(11):964--968,
  2020.

\bibitem{anderson1991infectious}
Roy~M Anderson and Robert~M May.
\newblock {\em Infectious diseases of humans: dynamics and control}.
\newblock Oxford university press, 1991.

\bibitem{Leung2017}
Kathy Leung, Mark Jit, Eric H.~Y. Lau, and Joseph~T. Wu.
\newblock {\em Scientific Reports}, 7(1), 08 2017.

\bibitem{mossong2008social}
Jo{\"e}l Mossong, Niel Hens, Mark Jit, Philippe Beutels, Kari Auranen, Rafael
  Mikolajczyk, Marco Massari, Stefania Salmaso, Gianpaolo~Scalia Tomba, Jacco
  Wallinga, et~al.
\newblock Social contacts and mixing patterns relevant to the spread of
  infectious diseases.
\newblock {\em PLoS medicine}, 5(3):e74, 2008.

\bibitem{melegaro2017}
Alessia Melegaro, Emanuele Del~Fava, Piero Poletti, Stefano Merler, Constance
  Nyamukapa, John Williams, Simon Gregson, and Piero Manfredi.
\newblock Social contact structures and time use patterns in the manicaland
  province of zimbabwe.
\newblock {\em PLOS ONE}, 12(1):1--17, 01 2017.

\bibitem{Mistry2021}
Dina Mistry, Maria Litvinova, Ana {Pastore y Piontti}, Matteo Chinazzi, Laura
  Fumanelli, Marcelo F~C Gomes, Syed~A Haque, Quan-Hui Liu, Kunpeng Mu, Xinyue
  Xiong, M~Elizabeth Halloran, Ira~M Longini, Stefano Merler, Marco Ajelli, and
  Alessandro Vespignani.
\newblock {Inferring high-resolution human mixing patterns for disease
  modeling}.
\newblock {\em Nature Communications}, 12(1):323, 2021.

\bibitem{prem2017projecting}
Kiesha Prem, Alex~R Cook, and Mark Jit.
\newblock Projecting social contact matrices in 152 countries using contact
  surveys and demographic data.
\newblock {\em PLoS computational biology}, 13(9):e1005697, 2017.

\bibitem{grijalva2015household}
Carlos~G Grijalva, Nele Goeyvaerts, Hector Verastegui, Kathryn~M Edwards, Ana~I
  Gil, Claudio~F Lanata, Niel Hens, et~al.
\newblock A household-based study of contact networks relevant for the spread
  of infectious diseases in the highlands of peru.
\newblock {\em PloS one}, 10(3):e0118457, 2015.

\bibitem{gozzi2022anatomy}
Nicol{\`o} Gozzi, Matteo Chinazzi, Jessica~T Davis, Kunpeng Mu, Ana Pastore~y
  Piontti, Marco Ajelli, Nicola Perra, and Alessandro Vespignani.
\newblock Anatomy of the first six months of covid-19 vaccination campaign in
  italy.
\newblock {\em PLoS Computational Biology}, 18(5):e1010146, 2022.

\bibitem{Zhang2020}
Juanjuan Zhang, Maria Litvinova, Yuxia Liang, Yan Wang, Wei Wang, Shanlu Zhao,
  Qianhui Wu, Stefano Merler, C{\'e}cile Viboud, Alessandro Vespignani, Marco
  Ajelli, and Hongjie Yu.
\newblock Changes in contact patterns shape the dynamics of the covid-19
  outbreak in china.
\newblock {\em Science}, 368(6498):1481--1486, 2020.

\bibitem{tizzoni2022addressing}
Michele Tizzoni, Elaine~O Nsoesie, Laetitia Gauvin, M{\'a}rton Karsai, Nicola
  Perra, and Shweta Bansal.
\newblock Addressing the socioeconomic divide in computational modeling for
  infectious diseases.
\newblock {\em Nature Communications}, 13(1):1--7, 2022.

\bibitem{buckee2021thinking}
Caroline Buckee, Abdisalan Noor, and Lisa Sattenspiel.
\newblock Thinking clearly about social aspects of infectious disease
  transmission.
\newblock {\em Nature}, 595(7866):205--213, 2021.

\bibitem{bedson2021review}
Jamie Bedson, Laura~A Skrip, Danielle Pedi, Sharon Abramowitz, Simone Carter,
  Mohamed~F Jalloh, Sebastian Funk, Nina Gobat, Tamara Giles-Vernick, Gerardo
  Chowell, et~al.
\newblock A review and agenda for integrated disease models including social
  and behavioural factors.
\newblock {\em Nature human behaviour}, 5(7):834--846, 2021.

\bibitem{zelner2022there}
Jon Zelner, Nina~B Masters, Ramya Naraharisetti, Sanyu~A Mojola, Merlin
  Chowkwanyun, and Ryan Malosh.
\newblock There are no equal opportunity infectors: epidemiological modelers
  must rethink our approach to inequality in infection risk.
\newblock {\em PLoS computational biology}, 18(2):e1009795, 2022.

\bibitem{karsai2020hungary}
M{\'a}rton Karsai, J{\'u}lia Koltai, Orsolya V{\'a}s{\'a}rhelyi, and Gergely
  R{\"o}st.
\newblock Hungary in mask/maszk in hungary.
\newblock {\em Corvinus Journal of Sociology and Social Policy}, (2), 2020.

\bibitem{koltai2022reconstructing}
J{\'u}lia Koltai, Orsolya V{\'a}s{\'a}rhelyi, Gergely R{\"o}st, and M{\'a}rton
  Karsai.
\newblock Reconstructing social mixing patterns via weighted contact matrices
  from online and representative surveys.
\newblock {\em Scientific Reports}, 12(1):1--12, 2022.

\bibitem{brankston2021quantifying}
Gabrielle Brankston, Eric Merkley, David~N Fisman, Ashleigh~R Tuite, Zvonimir
  Poljak, Peter~J Loewen, and Amy~L Greer.
\newblock Quantifying contact patterns in response to covid-19 public health
  measures in canada.
\newblock {\em BMC public health}, 21(1):1--10, 2021.

\bibitem{trentini2022investigating}
Filippo Trentini, Adriana Manna, Nicoletta Balbo, Valentina Marziano, Giorgio
  Guzzetta, Samantha O’Dell, Allisandra~G Kummer, Maria Litvinova, Stefano
  Merler, Marco Ajelli, et~al.
\newblock Investigating the relationship between interventions, contact
  patterns, and sars-cov-2 transmissibility.
\newblock {\em Epidemics}, 40:100601, 2022.

\bibitem{rohani}
Matt~J. Keeling and Pejman Rohani.
\newblock {\em Modeling Infectious Diseases in Humans and Animals}.
\newblock Princeton University Press, 2008.

\bibitem{hethcote2000mathematics}
Herbert~W Hethcote.
\newblock The mathematics of infectious diseases.
\newblock {\em SIAM review}, 42(4):599--653, 2000.

\bibitem{sandor2022covid}
J{\'a}nos S{\'a}ndor, Ferenc Vincze, Maya~Liza Shrikant, L{\'a}szl{\'o}
  K{\H{o}}r{\"o}si, L{\'a}szl{\'o} Ulicska, Karolina K{\'o}sa, and R{\'o}za
  {\'A}d{\'a}ny.
\newblock Covid-19 vaccination coverage in deprived populations living in
  segregated colonies: A nationwide cross-sectional study in hungary.
\newblock {\em Plos one}, 17(2):e0264363, 2022.

\bibitem{cadeddu2022planning}
Chiara Cadeddu, Aldo Rosano, Leonardo Villani, Giovanni~Battista Coiante,
  Ilaria Minicucci, Domenico Pascucci, and Chiara de~Waure.
\newblock Planning and organization of the covid-19 vaccination campaign: An
  overview of eight european countries.
\newblock {\em Vaccines}, 10(10):1631, 2022.

\bibitem{minter2019approximate}
Amanda Minter and Renata Retkute.
\newblock Approximate bayesian computation for infectious disease modelling.
\newblock {\em Epidemics}, 29:100368, 2019.

\bibitem{sunnaaker2013approximate}
Mikael Sunn{\aa}ker, Alberto~Giovanni Busetto, Elina Numminen, Jukka Corander,
  Matthieu Foll, and Christophe Dessimoz.
\newblock Approximate bayesian computation.
\newblock {\em PLoS computational biology}, 9(1):e1002803, 2013.

\bibitem{kimittud}
Covid-19 mortality and recovery data on settlement level, atlatszo.hu
\newblock Accessed: May 9, 2023.

\bibitem{oroszi2022characteristics}
Beatrix Oroszi, Attila Juh{\'a}sz, Csilla Nagy, Judit~Krisztina Horv{\'a}th,
  Krisztina~Eszter Koml{\'o}s, Gerg{\H{o}} T{\'u}ri, Martin McKee, and R{\'o}za
  {\'A}d{\'a}ny.
\newblock Characteristics of the third covid-19 pandemic wave with special
  focus on socioeconomic inequalities in morbidity, mortality and the uptake of
  covid-19 vaccination in hungary.
\newblock {\em Journal of personalized medicine}, 12(3):388, 2022.

\bibitem{oroszi2021unequal}
Beatrix Oroszi, Attila Juh{\'a}sz, Csilla Nagy, Judit~Krisztina Horv{\'a}th,
  Martin McKee, and R{\'o}za {\'A}d{\'a}ny.
\newblock Unequal burden of covid-19 in hungary: a geographical and
  socioeconomic analysis of the second wave of the pandemic.
\newblock {\em BMJ global health}, 6(9):e006427, 2021.

\bibitem{manna2023generalized}
Adriana Manna, Lorenzo Dall'Amico, Michele Tizzoni, Marton Karsai, and Nicola
  Perra.
\newblock Generalized contact matrices for epidemic modeling.
\newblock {\em arXiv preprint arXiv:2306.17250}, 2023.

\bibitem{naih}
Nemzeti adatvédelmi és információ szabadság hatóság, date of access
  2023.05.23.

\bibitem{ProxyContactDef}
Surveillance definitions for covid-19, european centre for disease prevention
  and control, date of access 2023.05.23.

\bibitem{feehan2021quantifying}
Dennis~M Feehan and Ayesha~S Mahmud.
\newblock Quantifying population contact patterns in the united states during
  the covid-19 pandemic.
\newblock {\em Nature communications}, 12(1):1--9, 2021.

\bibitem{brambor2006understanding}
Thomas Brambor, William~Roberts Clark, and Matt Golder.
\newblock Understanding interaction models: Improving empirical analyses.
\newblock {\em Political analysis}, 14(1):63--82, 2006.

\bibitem{mood2010logistic}
Carina Mood.
\newblock Logistic regression: Why we cannot do what we think we can do, and
  what we can do about it.
\newblock {\em European sociological review}, 26(1):67--82, 2010.

\bibitem{allison1999comparing}
Paul~D Allison.
\newblock Comparing logit and probit coefficients across groups.
\newblock {\em Sociological methods \& research}, 28(2):186--208, 1999.

\bibitem{kissler2020projecting}
Stephen~M Kissler, Christine Tedijanto, Edward Goldstein, Yonatan~H Grad, and
  Marc Lipsitch.
\newblock Projecting the transmission dynamics of sars-cov-2 through the
  postpandemic period.
\newblock {\em Science}, 368(6493):860--868, 2020.

\bibitem{backer2020incubation}
Jantien~A Backer, Don Klinkenberg, and Jacco Wallinga.
\newblock Incubation period of 2019 novel coronavirus (2019-ncov) infections
  among travellers from wuhan, china, 20--28 january 2020.
\newblock {\em Eurosurveillance}, 25(5):2000062, 2020.

\bibitem{blackwood2018introduction}
Julie~C Blackwood and Lauren~M Childs.
\newblock An introduction to compartmental modeling for the budding infectious
  disease modeler.
\newblock 2018.

\bibitem{voko2022nationwide}
Zolt{\'a}n Vok{\'o}, Zolt{\'a}n Kiss, Gy{\"o}rgy Surj{\'a}n, Orsolya
  Surj{\'a}n, Zs{\'o}fia Barcza, Bernadett P{\'a}lyi, Eszter Formanek-Balku,
  Gerg{\H{o}}~Attila Moln{\'a}r, R{\'o}bert Herczeg, Attila Gyenesei, et~al.
\newblock Nationwide effectiveness of five sars-cov-2 vaccines in hungary—the
  hun-ve study.
\newblock {\em Clinical Microbiology and Infection}, 28(3):398--404, 2022.

\bibitem{shapiro2021efficacy}
Julia Shapiro, Natalie~E Dean, Zachary~J Madewell, Yang Yang, M~Elizabeth
  Halloran, and Ira Longini.
\newblock Efficacy estimates for various covid-19 vaccines: what we know from
  the literature and reports.
\newblock {\em MedRxiv}, pages 2021--05, 2021.

\bibitem{statista}
Statista.
\newblock Hungary: Number of people vaccinated against covid-19.
\newblock
  \url{https://www.statista.com/statistics/1196109/hungary-number-of-people-vaccinated-against-covid-19/},
  2023.
\newblock Accessed: May 9, 2023.

\bibitem{salje2020estimating}
Henrik Salje, C{\'e}cile Tran~Kiem, No{\'e}mie Lefrancq, No{\'e}mie Courtejoie,
  Paolo Bosetti, Juliette Paireau, Alessio Andronico, Nathana{\"e}l Hoz{\'e},
  Jehanne Richet, Claire-Lise Dubost, et~al.
\newblock Estimating the burden of sars-cov-2 in france.
\newblock {\em Science}, 369(6500):208--211, 2020.

\end{thebibliography}

\end{document}